\documentclass[showkeys,nofootinbib]{revtex4}

\usepackage{amsfonts}
\usepackage{amssymb}
\usepackage{natbib}
\usepackage[latin1]{inputenc}
\usepackage[babel]{csquotes}
\usepackage{graphicx}
\usepackage[english]{babel}
\usepackage{multirow}

\usepackage[T1]{fontenc}
\usepackage[centertags]{amsmath}

\usepackage{mathrsfs}
\usepackage{amsmath}
\usepackage{amsthm}
\usepackage{amssymb}
\usepackage{mathtools}
\usepackage{wasysym}
\usepackage{subfigure}
\DeclareFontFamily{U}{mathb}{\hyphenchar\font45}
\DeclareFontShape{U}{mathb}{m}{n}{
      <5> <6> <7> <8> <9> <10> gen * mathb
      <10.95> mathb10 <12> <14.4> <17.28> <20.74> <24.88> mathb12
      }{}
\DeclareSymbolFont{mathb}{U}{mathb}{m}{n}

\DeclareMathSymbol{\Sun}{3}{mathb}{"40}

\begin{document}
\title{Including topology change in Loop Quantum Gravity with topspin network formalism with application to homogeneous and isotropic cosmology}
\author{Mattia Villani}
\affiliation{DISPEA, Universit\`a di Urbino ``Carlo Bo'' via Santa Chiara, 27 61029, Urbino, Italy}
\affiliation{INFN - Sezione di Firenze via B.Rossi, 1 50019, Sesto Fiorentino, Florence, Italy}
\email{mattia.villani@uniurb.it}
\date{\today}
\begin{abstract}
We apply topspin network formalism to Loop Quantum Gravity in order to include in the theory the possibility of changes in the topology of spacetime. We apply this formalism to three toy models: with the first, we find that the topology can actually change due to the action of the Hamiltonian constraint and with the second we find that the final state might be a superposition of states with different topologies. In the third and last application, we consider an homogeneous and isotropic Universe, calculating the difference equation that describes the evolution of the system and which are the final topological states after the action of the Hamiltonian constraint. For this last case, we  also calculate the transition amplitudes and probabilities from the initial to the final states.
\end{abstract}
\maketitle

\section{Introduction}

One of the biggest problems of modern physics is the difficult relation between general relativity and quantum mechanics. They describe fairly well the physical world at different scales: quantum mechanics describes the microscopical behavior of matter and its components, while general relativity, on the other hand, describes the physics of the universe as a whole and of compact objects like black holes and neutron stars, where gravity is strong. Both are confirmed by a number of experimental works, but even though general relativity is an extremely successful theory, it has a flaw: it predicts the existence of singularities, regions of spacetime where the curvature is infinite. This is interpreted as a sign that the physics quest to the description of the world is not complete: a new theory is needed, a quantum theory of gravity that brings together general relativity and quantum mechanics. However, quantum mechanics implies a very different view of the world which is hard to reconcile with that of general relativity. For example, the former is formulated in terms of \emph{systems} limited in space, but infinitely extended in time, while the latter describes the world in terms of \emph{events} limited both in time and in space \cite{LM}. Other problems are the non-locality of the superposition principle and the locality implied by the equivalence principle and the entanglement \cite{bau}. On top of that, in quantum mechanics spacetime is flat and governed by global global symmetries such as the Poincarr\'e group while in general relativity  spacetime is curved and symmetries are not global \cite{rov}. Physicists are exploring many roads in their quest to a quantum gravity theory, among them we cite string theory \cite{st1} and loop quantum gravity (LQG) (see \cite{rev1,rov,lrr} for reviews and \cite{rev2} for a complete mathematical introduction on the matter). 

String theory approach to quantum gravity is perturbative: it fixes a background flat Minkowski metric that gives the causal structure and defines on it all other fields. On the other hand, LQG approach is canonical: it starts with the constrained Hamiltonian of general relativity and classical Poisson brackets between Ashtekar variables \cite{ash} and quantizes them following Dirac's approach \cite{rev2,dir,lect}. Moreover, the theory is background independent, i.e. the theory does not fix a background metric for the spacetime on which the dynamics takes place, but instead derives it from the equations of motion. However this is not the most general approach, since the theory does fix the topology of the spacetime. The topology, however, might be dynamical at quantum level, so it is important to find ways in which the topology of the spacetime could be derived dynamically from the equations of motion. Many proposal have been made in order to include dynamically the topology in the theory, see for example \cite{topo1,topo2,topo3,tsn}; we focus on the approach described in \cite{tsn} based on branched coverings of 3-manifolds and the addition of a topological label on spin networks edges.

A strong topology theorem due to Alexander, \cite{tsn,ale}, demonstrates that every compact, oriented 3-manifold can be described as a branched covering over a knot embedded in the 3-sphere $S^3$. A branched cover is a submersion $p: M\mapsto S^3$ such that the restriction to the complement of a knot $\Gamma$ (the branch locus) embedded  in $S^3$ is an ordinary covering of degree $n$; this is indicated as $p_|:M-p^{-1}(\Gamma) \mapsto S^3-\Gamma$. A similar result also applies to compact connected 4-manifolds and the 4-sphere $S^4$, but in this case the branch locus is a 2-complex \cite{pier}. A refinement of the Alexander's theorem is due to Hilden and Montesinos \cite{H,M} who proved that every closed, oriented 3-manifold can be described as a 3-fold branched covering over a link in $S^3$. Yet another version of this theorem is due to Izmestiev and Joswig \cite{IJ} and states that a closed, oriented 3-manifold is described by a branched covering over a triangulation of $S^3$. A link between the second and the third form of the theorem is provided in \cite{but1,but2}, where the authors construct a triangulation of $S^3$ starting from a link embedded in the hyper-sphere. The branch locus $\Gamma$ completely describes the branched covering and it can be obtained starting from a triangulation of the manifold $M$ using the procedure described in \cite{knots}.

In \cite{tsn}, authors propose to use Alexander theorem to describe a manifold as a branched covering over $S^3$, and to add a topological color to the edges of a spin network obtaining what they call a \emph{topspin network}. This new color is selected from a representation of the fundamental group $\pi_1(S^3-\Gamma)$ of the unbranched space into the symmetric group $S_n$; heuristically, this color describes how to stitch together over the branched locus the various branches of the cover. A similar approach might be used to derive a \emph{topspin foam}, a spin foam on which topological label have been added. This is achieved with the notion of branched cover cobordism between two 3-manifolds (see \cite{tsn}).\footnote{\emph{Cobordism} means that the union of the two manifolds constitutes the boundary of a 4-manifold.}

Since with this new approach the topology of the manifold can change during evolution of the system, it is important to have methods to characterize topologically the manifold. We follow \cite{dust2} and use Fox free calculus \cite{fox1,fox2,fox3,fox4,fox5,fox6,fox7,comb}. With this method, one can calculate the fundamental group of the manifold which is a topological invariant and can be used to classify 3-manifolds \cite{class1,class2}. If it changes during the evolution of the system it means that the topology of the manifold has changed. As we shall see, it is also possible that a system is found in a superposition of states with different topologies: in this typically quantum situation, a probability is associated to the various states and there will be transition amplitudes from one state to another.

The plan of the paper is as follows: in Section \ref{sec:tsn} we review spin and topspin network and their use in quantum gravity; in Section \ref{sec:ham} we show how  the Hamiltonian constraint acts on topspin network;  in Section \ref{sec:appl} we apply the formalism to three simple models: the first two are intended to prove that the topology of the manifold can change due to the action of the Hamiltonian constraint and that the final state can be a superposition of states with different topologies; finally we apply the formalism of topspin network to a model of homogeneous and isotropic Universe and also calculate the transition amplitudes between the various final states; in Section \ref{sec:concl} we conclude our exposition.

\section{Spin and topspin networks}
\label{sec:tsn}

In LQG, the state of a system is described by a spin network which can be defined as a triple $(\Gamma,\rho,\iota)$ where $\Gamma$ is a graph embedded in the manifold $M$, $\rho$ is a coloring for the edges of $\Gamma$ taken by the representation $\rho_e$ of the group $SU(2)$ and $\iota$ is a labeling of the vertices $v$ of $\Gamma$ (called the intertwiners) given by:
\begin{equation*}
\iota: \rho_{e_1}\otimes\dots\otimes\rho_{e_n} \rightarrow \rho^\prime_{e_1}\otimes\dots\otimes\rho^\prime_{e_m}
\end{equation*}
where $\rho_{e_i}$ are the edges incoming in the vertex $v$ and $\rho^\prime_{e_i}$ are the edges outgoing from the vertex. 

Spin networks constitute a basis for the state of the system; these states are linearly independent and diagonalize operators that describe the 3-geometry of space such as the area and volume operators. Finally, spin networks provide a discrete picture of quantum geometry at Planck scale \cite{spinn}.

As stated in the introduction, in the new mathematical setting of topspin network, by applying the Alexander theorem and its later refinements, one views a 3-manifold as a branched covering of order $n$ over the 3-sphere $S^3$ branched over a graph $\Gamma$. In this way one can define a topspin network analogously to what is done for spin network \cite{tsn}, but adding an additional coloring $\sigma$ for the edges of the graph $\Gamma$ taken from a representation of the fundamental group of the complement of the graph into the symmetric group $S_n$, i.e. one considers the map $\sigma:\pi_1(S^3-\Gamma)\rightarrow S_n$; heuristically, this new coloring describes how to stitch together the various branches of the covering. The new topological colors $\sigma$ are not arbitrary but must satisfy the Wirtinger relations at vertices \cite{tsn}:
\begin{equation}
\prod_i \sigma_i =\prod_j\sigma_j
\end{equation}
where $i$ is an index running on edges incoming in the vertex and $j$ a label for the edges outgoing from the vertex. With this prescription, the topology is included in the description of the physical state through monodromies \cite{tsn}. It must be pointed out that the relation between 3-manifolds and branched covering is not bijective: a single manifold may be described by different branch-loci; authors in \cite{tsn} make the example of the homology sphere, which can be seen as a fivefold covering of $S^3$ over the trefoil knot, or as a threefold covering over the (2,5) torus knot. Which representation must be selected in this case of multiple possibilities depends on physical constraints, such as homogeneity and isotropy of the manifold of the system under consideration. 

Just like \emph{usual} spin networks, also topspin networks constitute a basis for the state of the system. It is possible to extend the usual Hilbert space used in LQG to the case of topspin networks by assuming that two topspin networks are orthogonal if they are not related by the covering moves described below and assuming that the inner product $\langle \psi, \psi^\prime\rangle$ of two topspin networks $\psi$ and $\psi^\prime$ that are equivalent by covering moves is defined \emph{as usual} forgetting about the topological coloring.

\begin{figure}[h]
\centering
\subfigure{\includegraphics[scale=0.35]{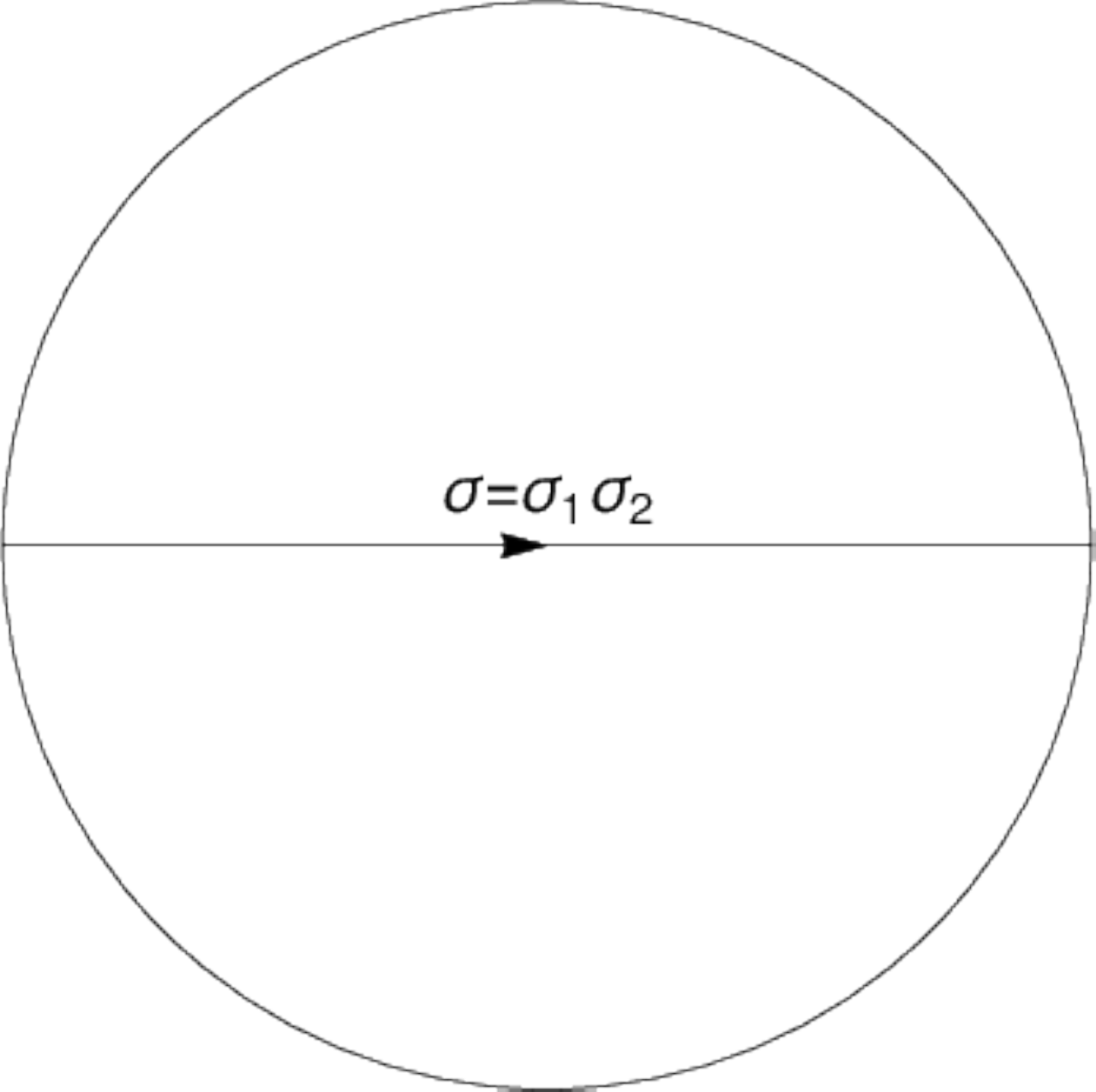}} \subfigure{\includegraphics[scale=0.35]{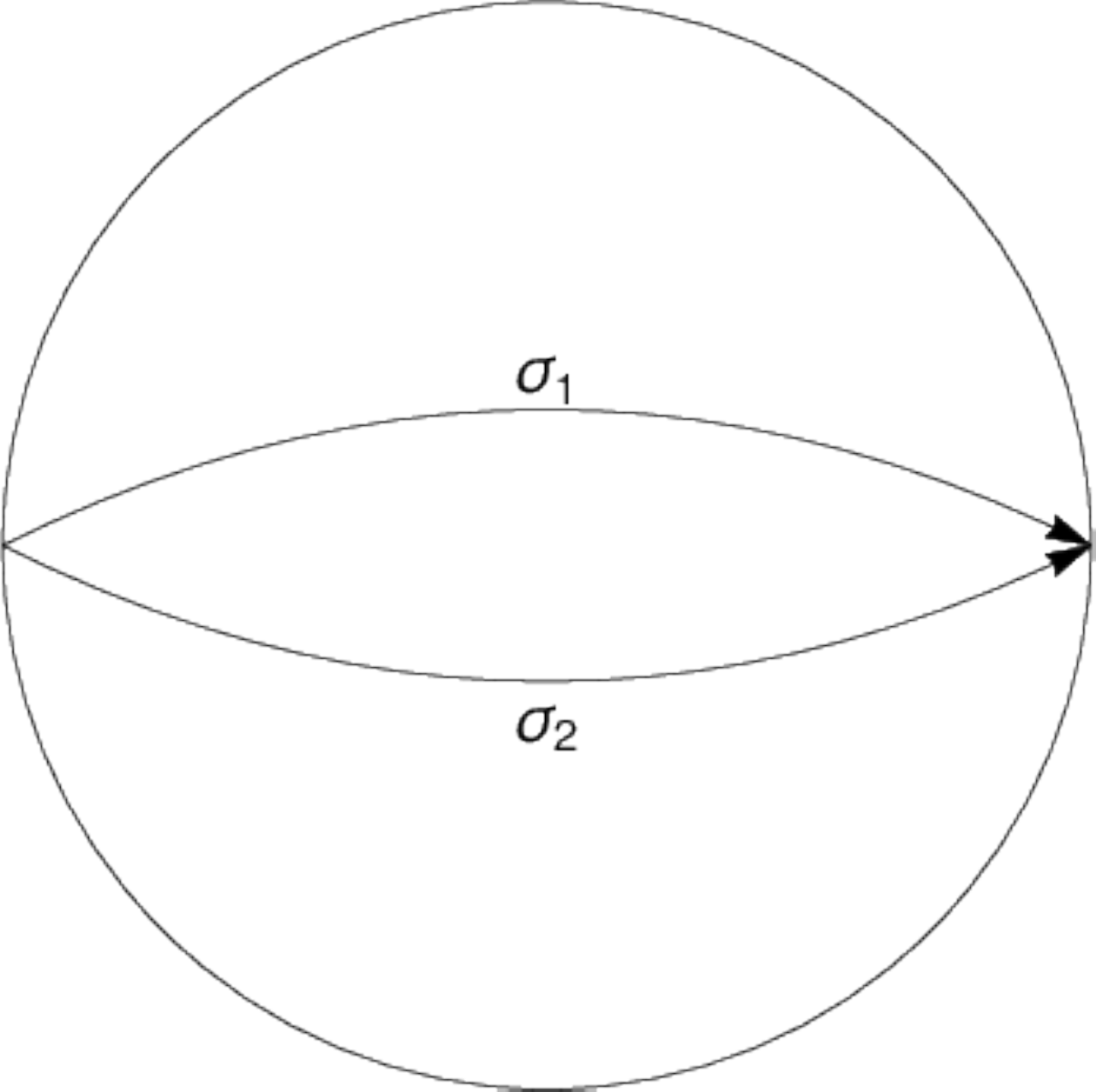}}
\caption{Vertex covering move: an edge with color $\sigma$ can be splitted into two edges with color $\sigma_1$ and $\sigma_2$ such that $\sigma=\sigma_1\sigma_2$.}\label{fig:cov_move_v1}
\end{figure}

\begin{figure}[h]
\centering
\subfigure{\includegraphics[scale=0.35]{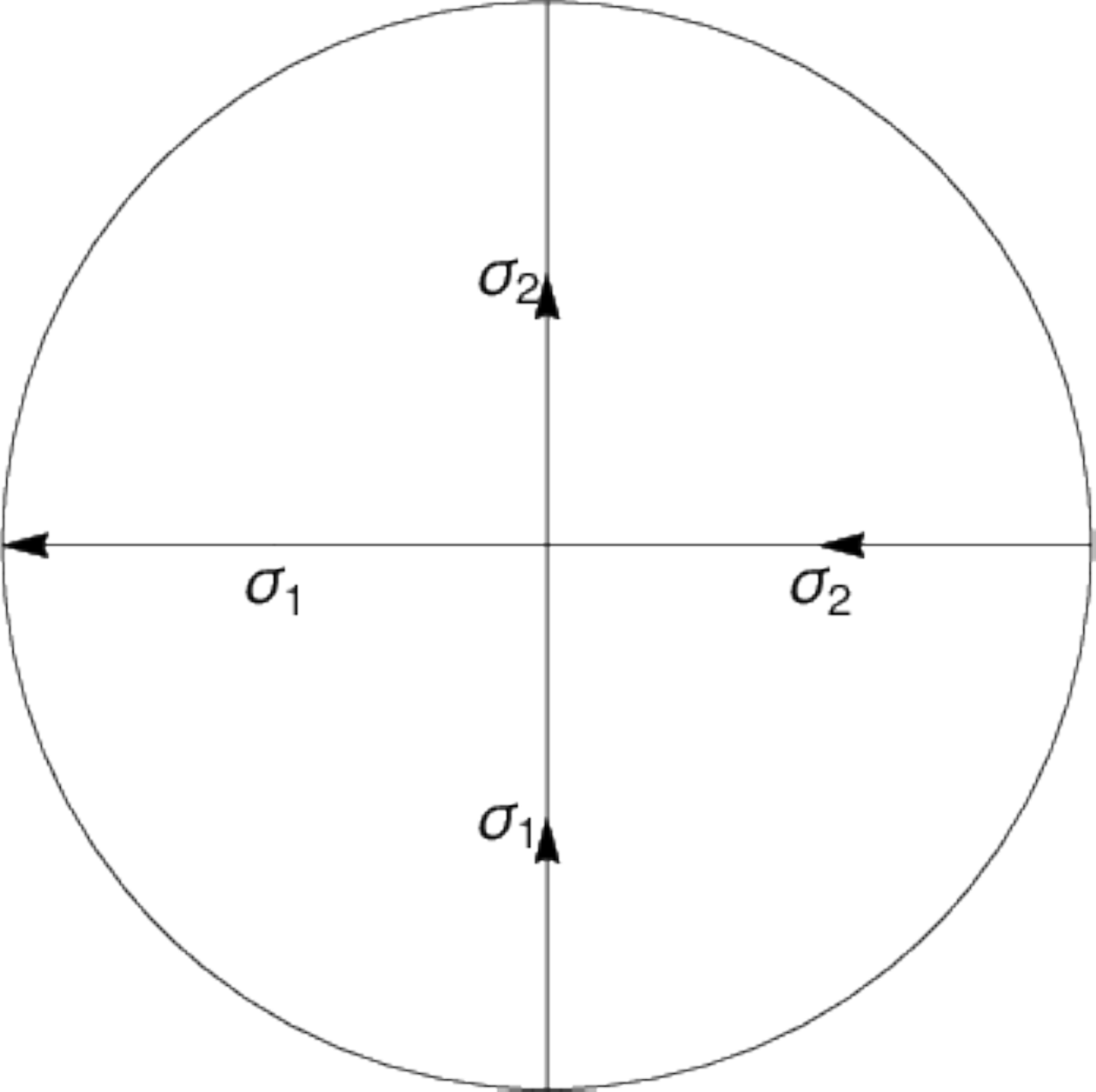}} \subfigure{\includegraphics[scale=0.35]{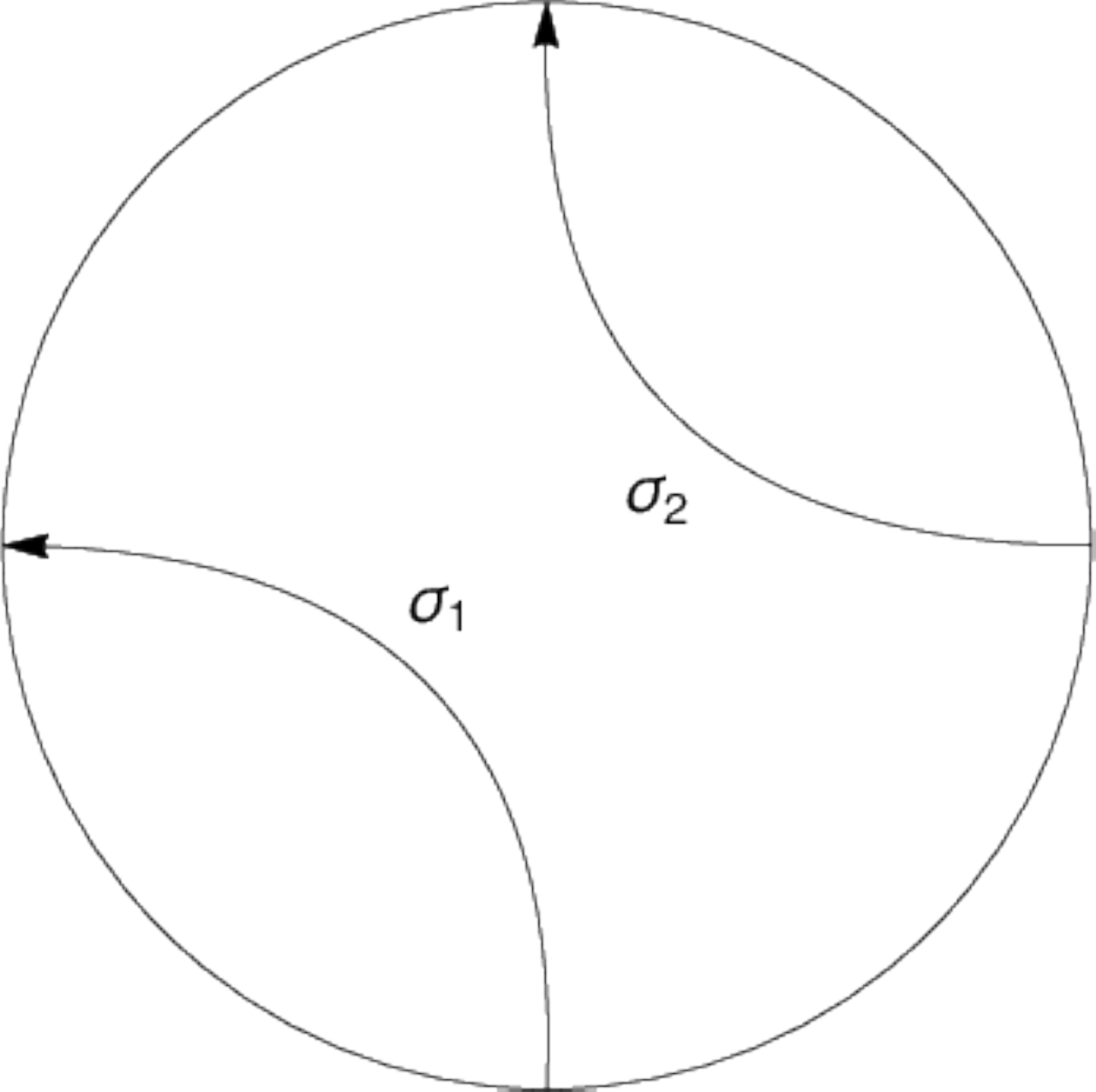}}
\caption{Vertex covering moves: a vertex can be eliminated if the incoming and the outgoing edges have the same coloring $\sigma_1$ and $\sigma_2$ (this trivially satisfies the Wirtinger relation in the left configuration).}\label{fig:cov_move_v2}
\end{figure}

\begin{figure}[h]
\centering
\subfigure{\includegraphics[scale=0.35]{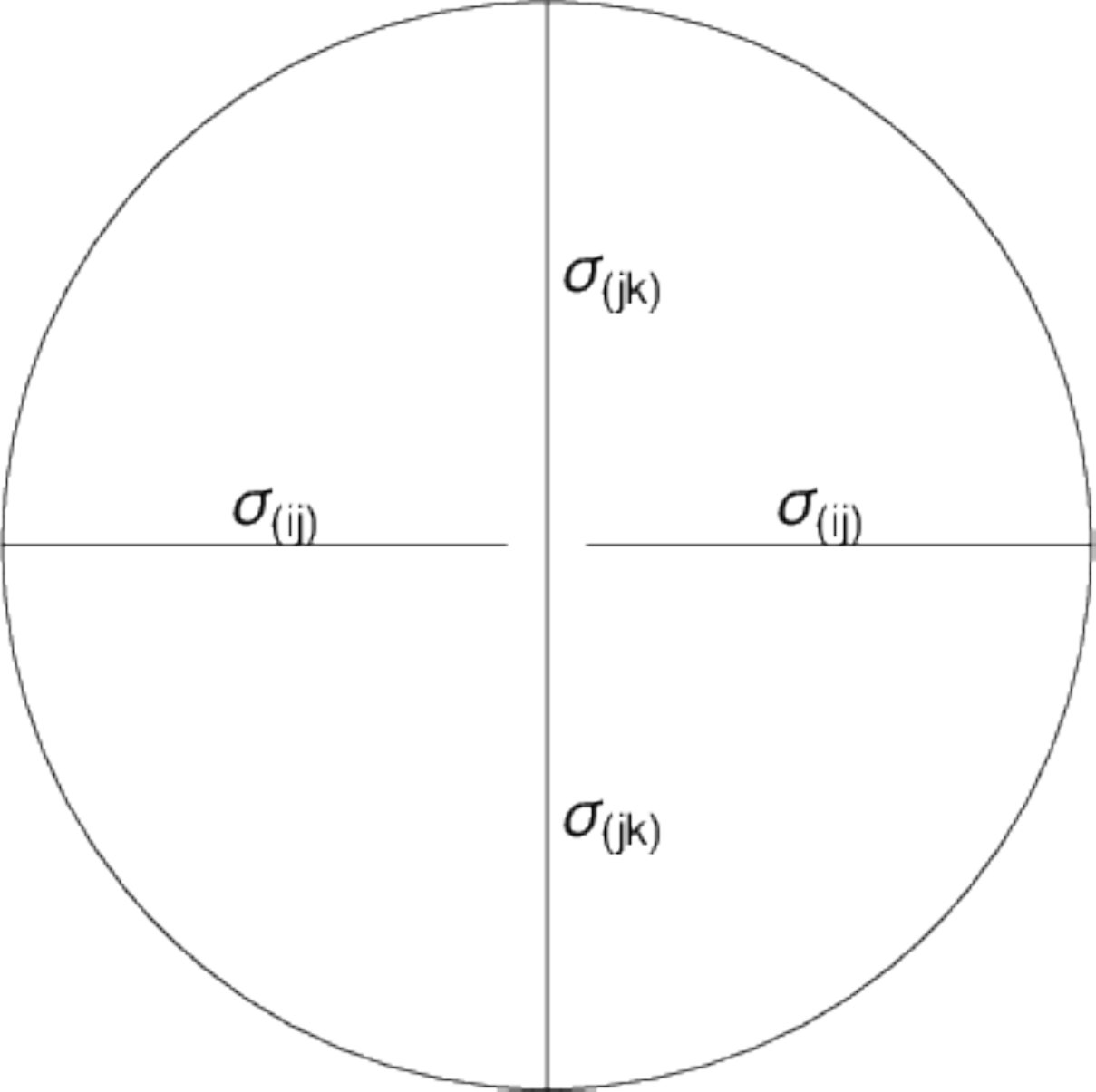}} \subfigure{\includegraphics[scale=0.35]{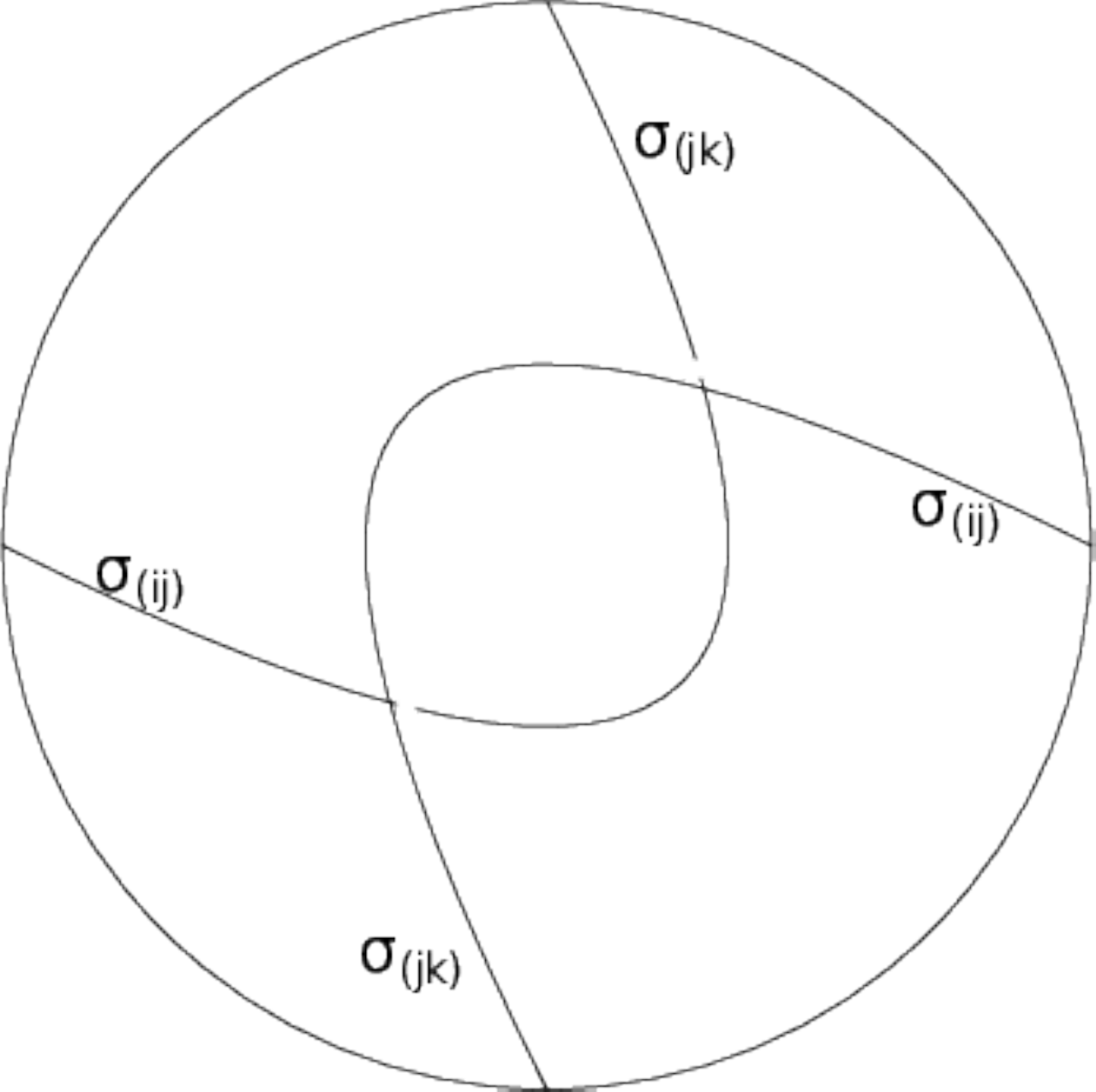}}
\caption{Crossing covering moves: edges with an over-crossing can be rerouted as in figure. An edge passing below another is indicated by two white stripes around the crossing.}\label{fig:cov_move_c1}
\end{figure}

\begin{figure}[h]
\centering
\subfigure{\includegraphics[scale=0.35]{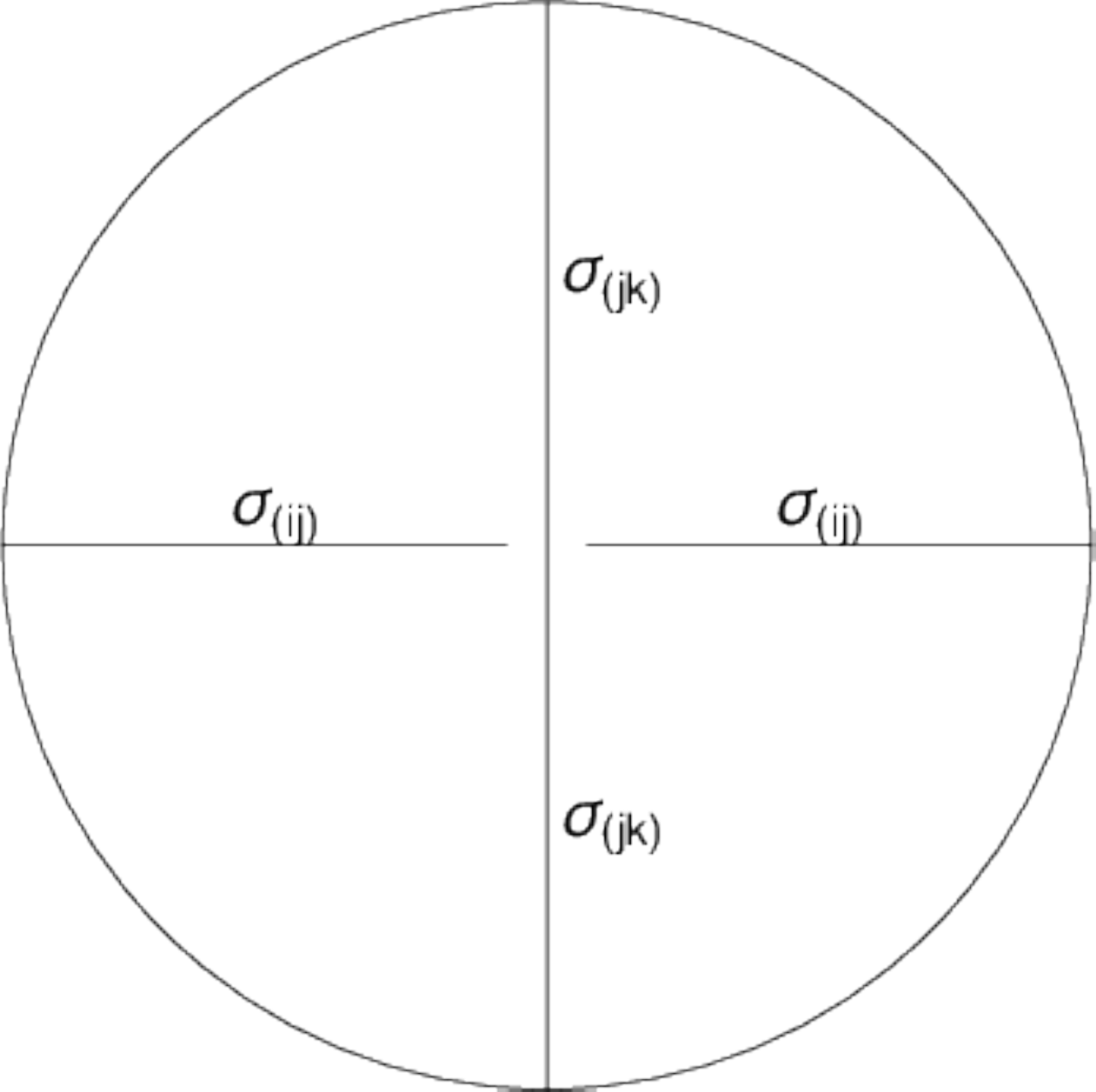}} \subfigure{\includegraphics[scale=0.35]{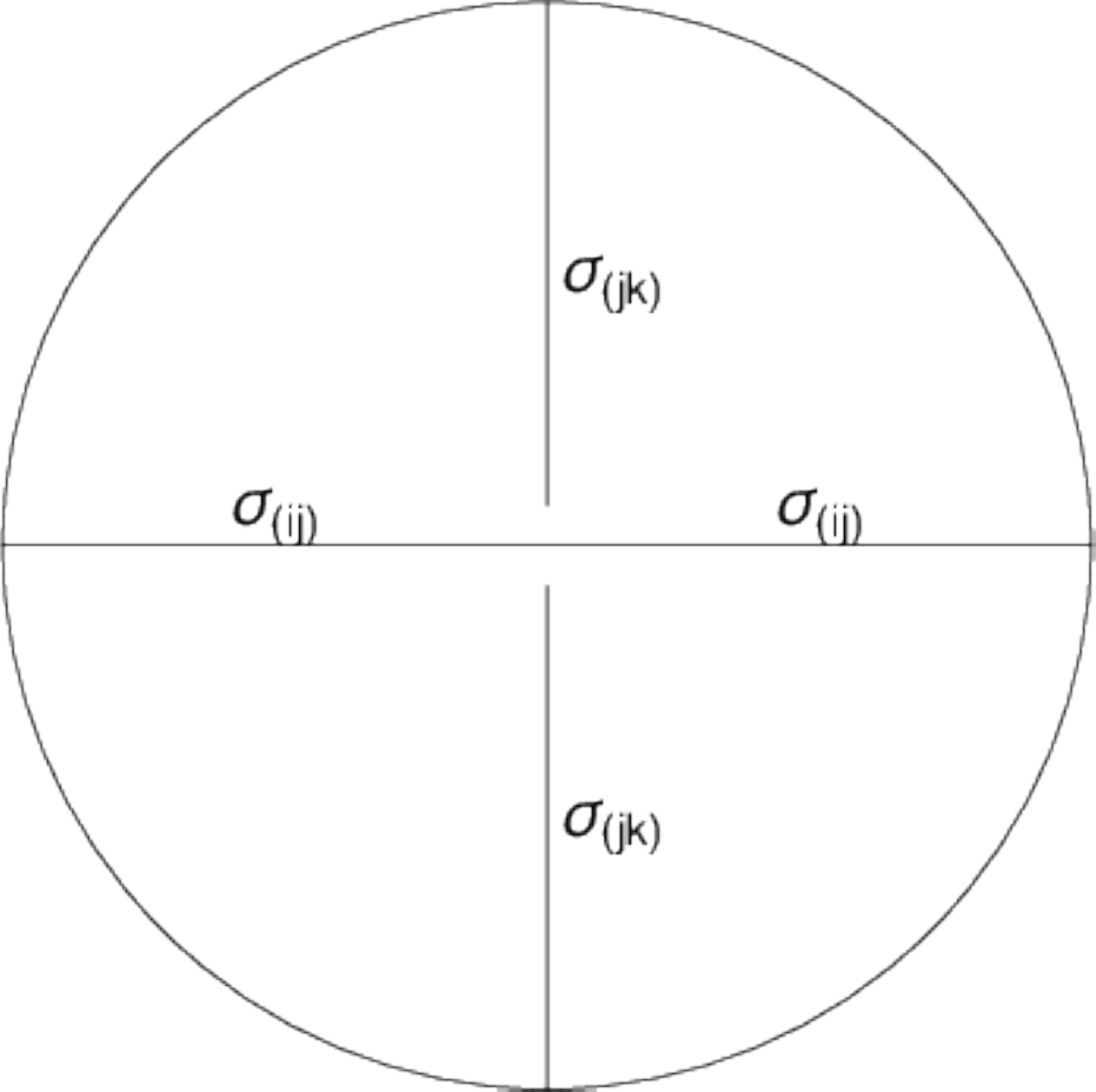}}
\caption{Crossing covering moves: an over-crossing can be changed to an under-crossing and vice versa. An edge passing below another is indicated by two white stripes around the crossing.}\label{fig:cov_move_c2}
\end{figure}

It is also possible that two different topspin networks (i.e. different branch-loci) describe the same physical state, i.e. the same manifold. The condition for this to happen is that one topspin network can be transformed into the other through covering and geometric moves \cite{tsn}. A covering move \cite{tsn} is a non isotopic modification of the labeled branch locus $\Gamma$ that results in another branch locus $\Gamma^\prime$ giving a different description of the same manifold M, i.e. the resulting manifolds are diffeomorphic to each other. There are two types of covering moves: at vertices and at crossings. Covering moves at vertices are themselves of two types and are reported in figures \ref{fig:cov_move_v1} and \ref{fig:cov_move_v2}: in figure \ref{fig:cov_move_v1} we show the case in which a representation $\sigma$ of $S_n$ is substituted by two representations such that $\sigma=\sigma_1\sigma_2$ (this splitting is by no means unique): the valence of the vertices is changed by this move, but the Wirtinger relations are satisfied; in figure \ref{fig:cov_move_v2} we see that a 4-vertices is removed: in this vertex there are two incoming edges colored as $\sigma_1$ and $\sigma_2$ and two outgoing edges also colored with $\sigma_1$ and $\sigma_2$: this move reduces by one the number of vertices and by two the number of edges, but the Wirtinger relations are again satisfied. Covering moves at crossings are also of two types: in  figure \ref{fig:cov_move_c1}, we see that an over crossing is changed by a redirection of the edges, while the move reported in figure \ref{fig:cov_move_c2} is simply a change in the over/down crossings of the two edges; both of these moves satisfy the Wirtinger relations. We notice that all the described moves are local, in fact they can be performed without affecting the graph outside a cell.

\begin{figure}[t]
\centering
\subfigure{\includegraphics[scale=0.35]{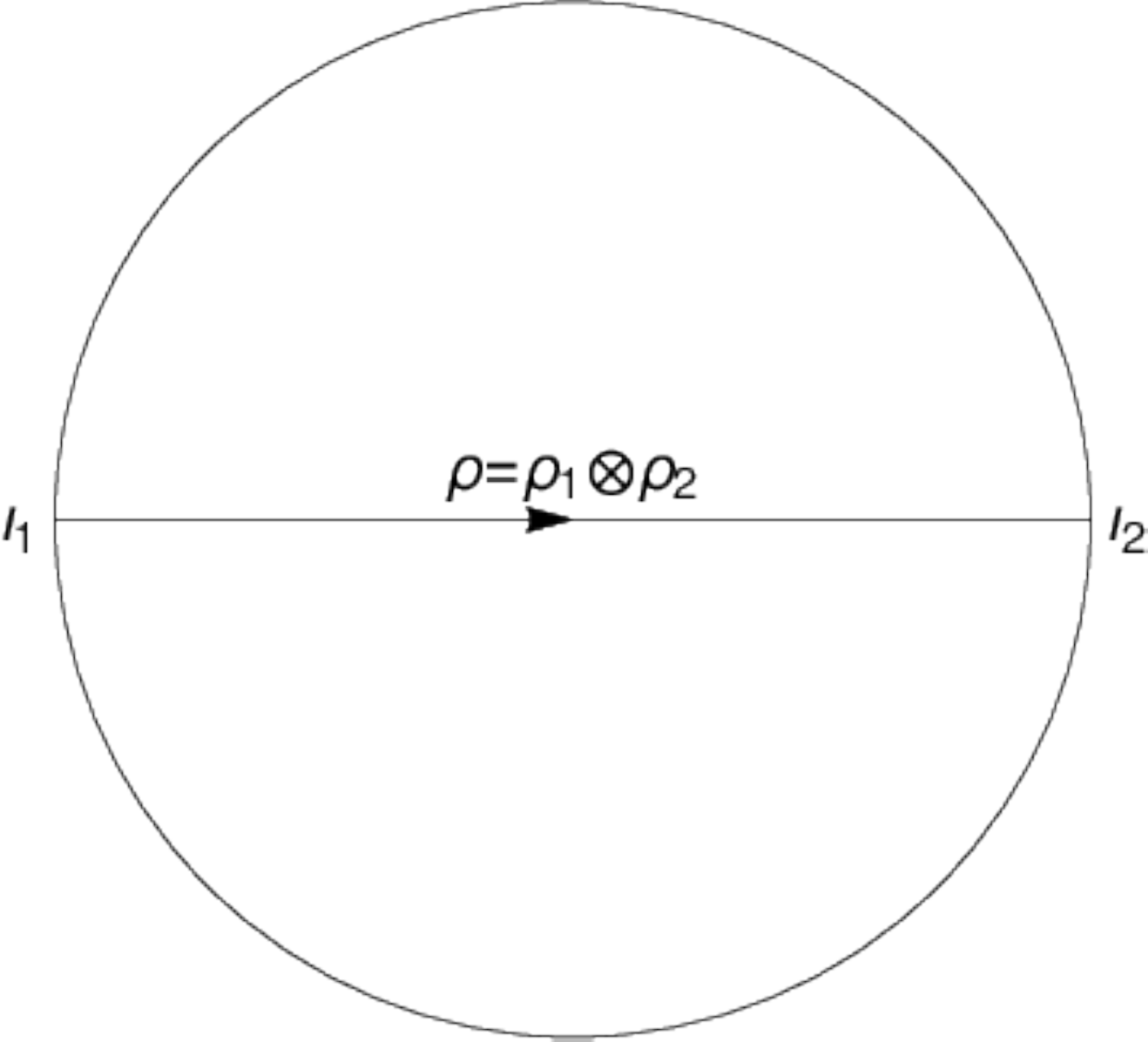}} \subfigure{\includegraphics[scale=0.35]{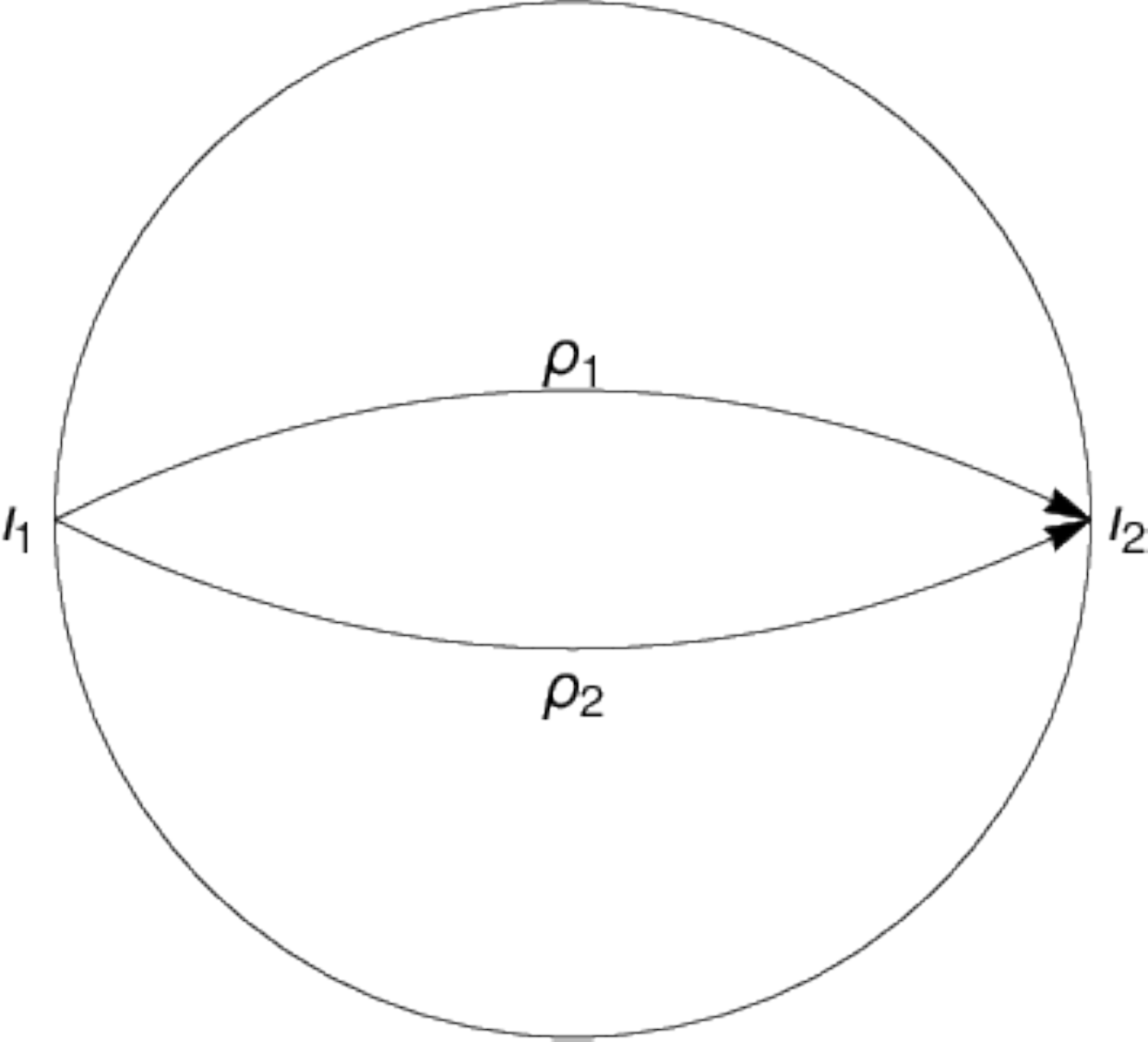}}
\caption{Geometric moves: an edge with geometric color $\rho=\rho_1\otimes\rho_2$ can be splitted into two edges with color $\rho_1$ and $\rho_2$.}\label{fig:geo1}
\end{figure}

\begin{figure}[t]
\centering
\subfigure{\includegraphics[scale=0.35]{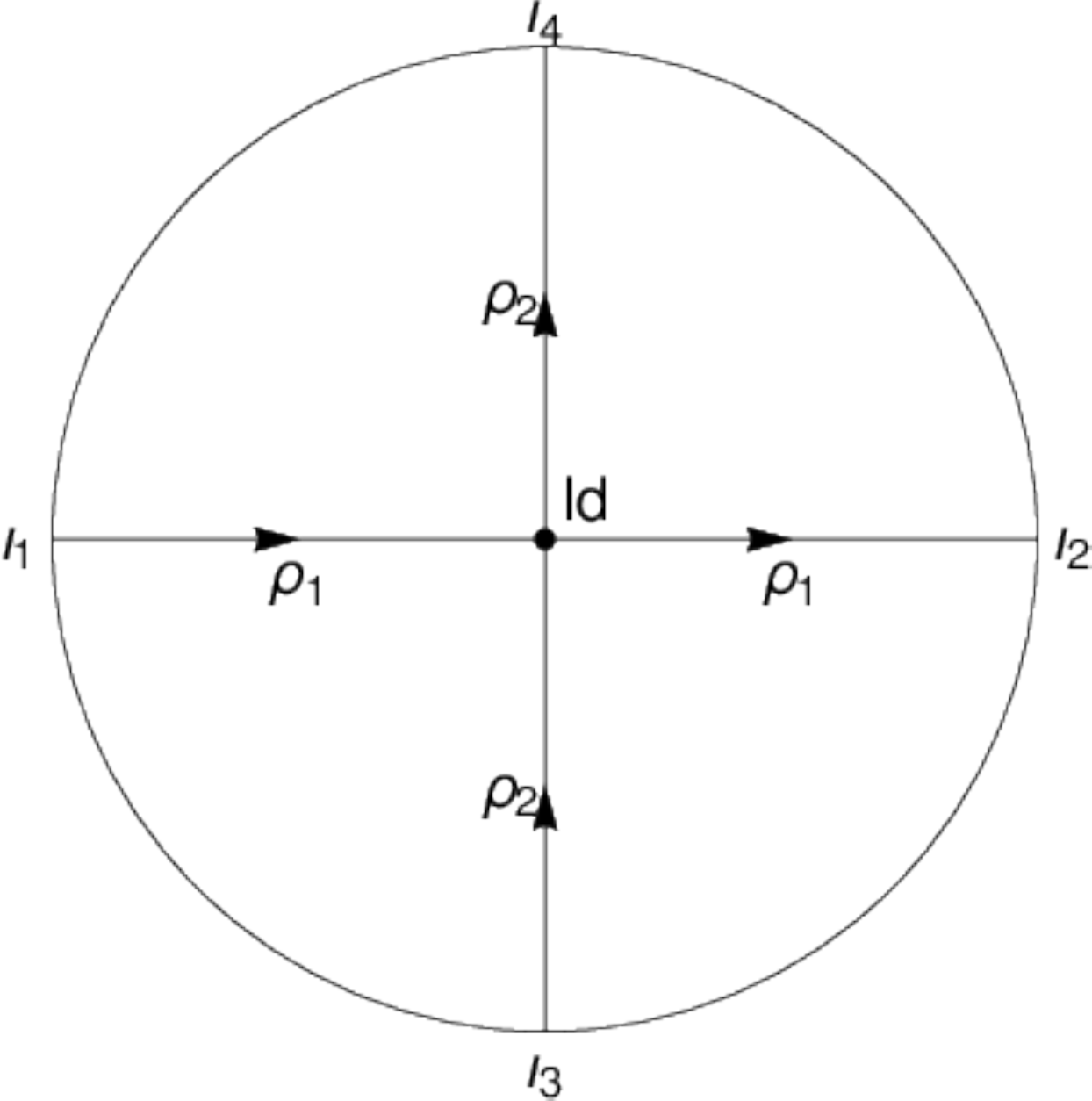}} \subfigure{\includegraphics[scale=0.35]{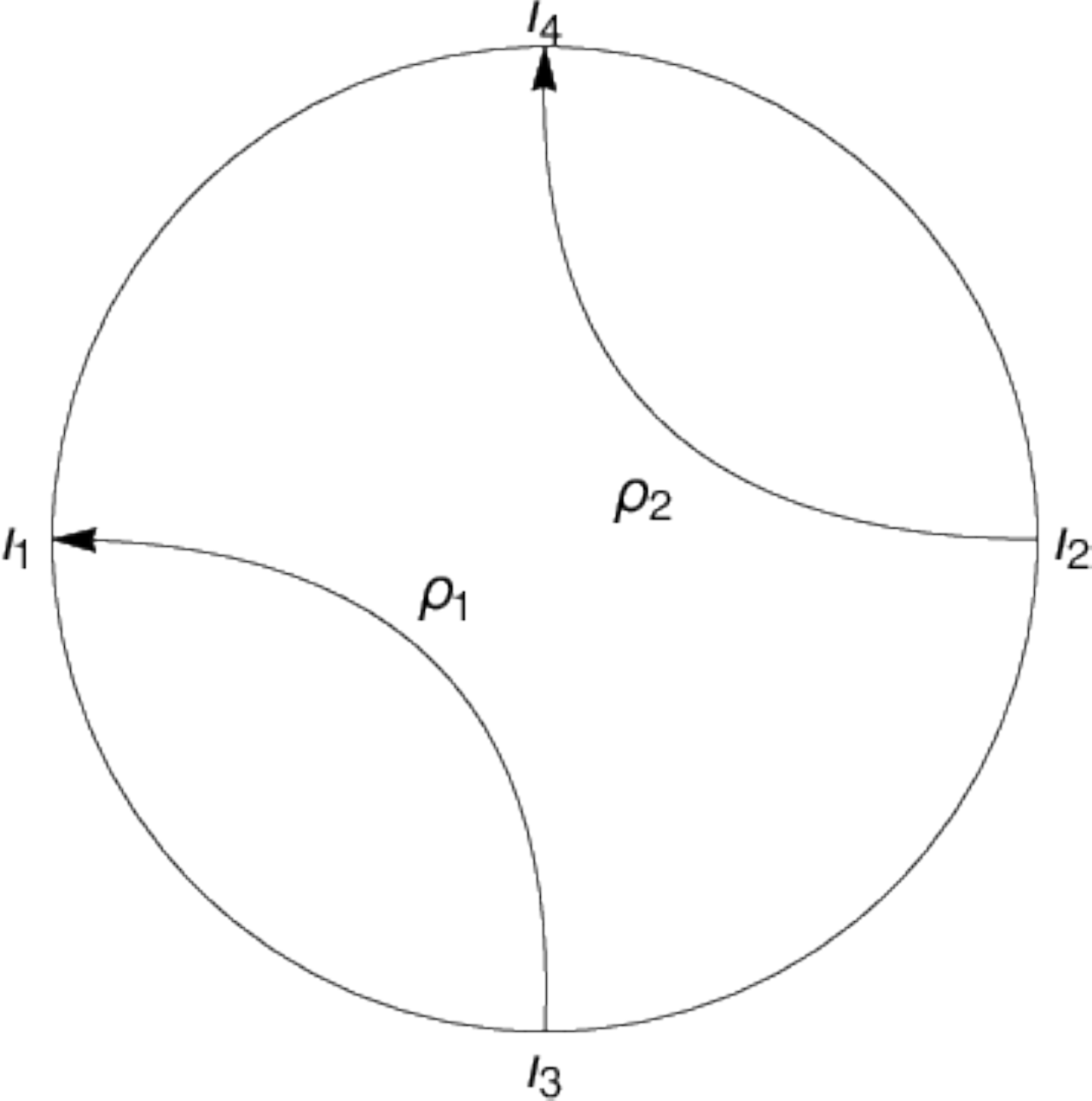}}
\caption{Geometric moves: the identity intertwiner can be removed.}\label{fig:geo2}
\end{figure}

With topspin networks, one must also ensure that topological and geometrical labeling are compatible, i.e. one must ask: if the topological coloring is changed with the above covering moves how should the geometrical labeling change in order to produces a manifold with the same geometry as the initial one? The answer to this question reduces to the two moves reported in figures \ref{fig:geo1} and \ref{fig:geo2}: in figure \ref{fig:geo1} we show a splitting of an edge like that of figure \ref{fig:cov_move_v1} into two that satisfy the relation $\rho=\rho_1\otimes\rho_2$ (as above, this splitting is by no means unique), while in figure \ref{fig:geo2} we show that a vertex with trivial intertwiner can be removed like in figure \ref{fig:cov_move_v2}, see also \cite{tsn}. 

Often, in LQG one considers graphs coming from triangulations of the manifold; in this new setting with topspin network, one can take a triangulation of $S^3$, $\Gamma^\prime$, which contains as subgraph the branch locus $\Gamma$ and assigns trivial topological coloring to the edges that are not part of $\Gamma$ (the new edges can carry, however, non-trivial geometrical coloring). In LQG it is also common to refine a triangulation and the graph, for example by barycentral subdivision; this can be done with topspin network too, in fact one may prescribe that the original edges keep their original coloring, while the new edges carry trivial topological color. In this way, all the arguments usually applied in LQG for nested graphs and directed limits carry over to the case of topspin network.





\section{Action of the Hamiltonian constraint on topspin networks}
\label{sec:ham}

\begin{figure}
\centering
\subfigure{\includegraphics[scale=0.5]{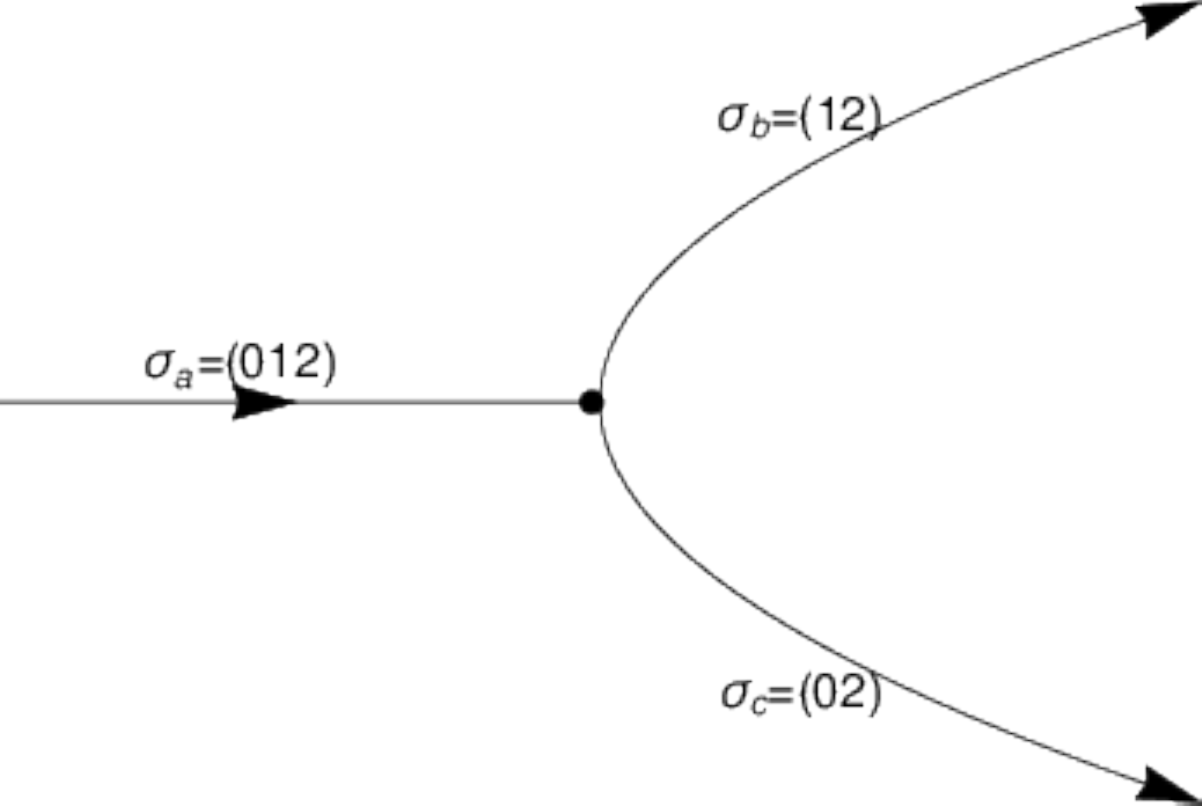}} \subfigure{\includegraphics[scale=0.5]{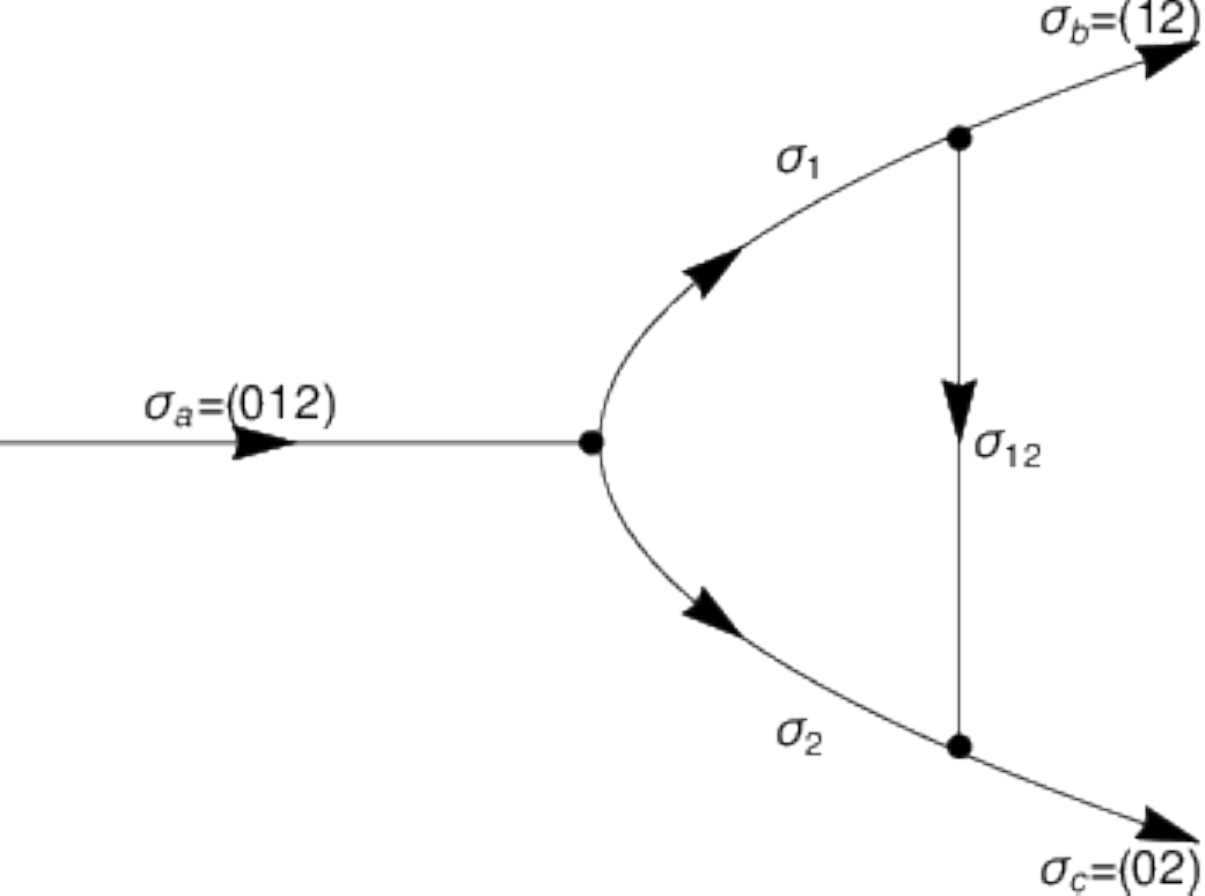}}
\caption{\emph{Left}: the initial graph. \emph{Right}: Graph obtained by the action of the Hamiltonian constraint; the extraordinary edge is the one labeled $\sigma_{12}$. Wirtinger relations imposes strong constraints on the topological colors and in this case uniquely fixes them (see main text).}\label{fig:tre}
\end{figure}

\begin{figure}
\centering
\includegraphics[scale=0.5]{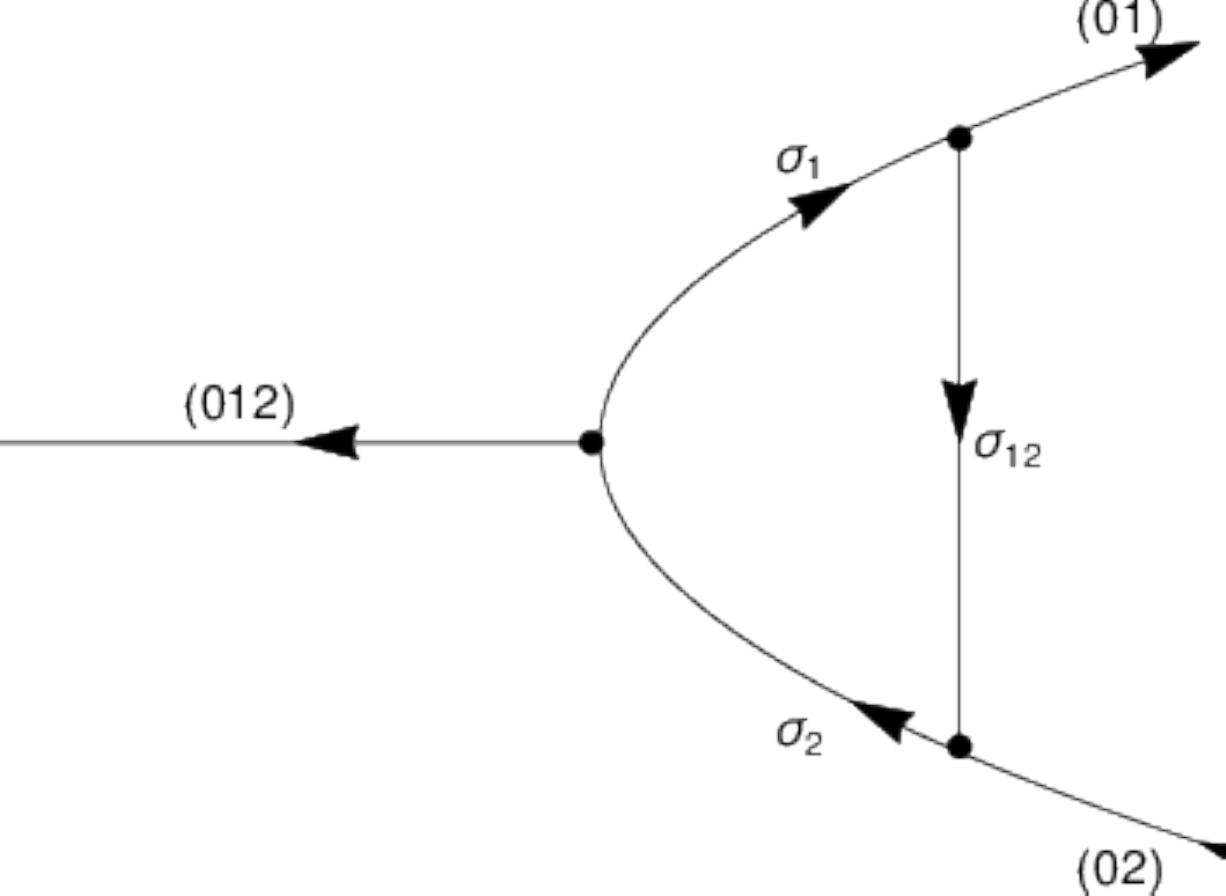}
\caption{Situation similar to figure \ref{fig:tre}, but in this case there are many possible solutions to the Wirtinger relations (see main text and table \ref{tab:poss}).}\label{fig:doppio}
\end{figure}

The action of the Hamiltonian constraint on spin networks results in the addition of a new edge called the extraordinary edge \cite{T1,T2}; this new edge is colored with a $j=1/2$ label from the fundamental $SU(2)$ representation and the $SU(2)$ coloring of neighboring edges is also modified consistently. The action on a topspin network is analogous, but now the extraordinary edge must carry a color describing the particular topological state and the neighboring edges must carry compatible topological colors. The topological label cannot be freely chosen, but must respect the Wirtinger relations at the vertices. When a new edge is added, there are three Wirtinger relations to be satisfied, in fact looking at the left panel of figure \ref{fig:tre}, if the initial edges carry indices $\sigma_a,\sigma_b,\sigma_c$ satisfying $\sigma_a=\sigma_b\sigma_c$ and the coloring of the new edges created by the action of the Hamiltonian constraint are $\sigma_1,\sigma_2,\sigma_{12}$, these new colors must satisfy the three relations:
\begin{equation}
\sigma_1\sigma_2=\sigma_a \qquad \sigma_1=\sigma_{12}\sigma_b \qquad \sigma_2\sigma_{12}=\sigma_c
\end{equation}
These relations constrain the possible values of the three $\sigma_i$'s. For example, consider the right panel of figure \ref{fig:tre}: assuming that the covering has 3 branches, the group $S_n$ that must be used to set the topological colors has three letters and it can be checked that the new edges must be colored as follows in order to respect the Wirtinger relations:
\begin{equation}
\sigma_1=(012) \qquad \sigma_2=(12) \qquad \sigma_{12}=(02)
\end{equation}
In this example there is a unique possible solution due to the particular orientation of the edges, any other configuration would violate the Wirtinger relations. On the other hand, in the situation of figure \ref{fig:doppio} there are six different possible solutions, listed in table \ref{tab:poss}. The fact that in this last case there are many possibile solutions is due to the different orientations of the edges in the graphs of figures \ref{fig:tre} and \ref{fig:doppio} which gives different Wirtinger relations. Each of these coloring must be considered in the process of determining the topology of the system; as we shall see in Section \ref{sec:super}, this might lead to a superposition of topologies in the final state.

\begin{table}
\centering
\begin{tabular}{ccc|ccc}
$\sigma_1$ & $\sigma_2$ & $\sigma_{12}$&$\sigma_1$ & $\sigma_2$ & $\sigma_{12}$\\
\hline
(02)  & (01) & id & (12) & (02) & (021) \\
(021) & (012) & (12) & (01) & (12) & (012)\\
(012) & id & (01) & id & (021) & (02)
\end{tabular}
\caption{Possible configurations of the colors for the graph of figure \ref{fig:doppio}. \emph{id} means \emph{identity}.}\label{tab:poss}
\end{table}

The addition of the new edge changes the relations between the elements of the fundamental group $\pi_1(S^3-\Gamma)$ of the unbranched covering and it can happen that the final topological state, i.e. the final fundamental group, is different from the initial one; we shall show a simple example of this in the applications section (see section \ref{sec:change}). This new state is characterized by a different fundamental group and also other topological invariants might change.



\section{Applications}
\label{sec:appl}

In this section we consider three applications: in the first, we show that the fundamental group of the final manifold might be different from the initial one; in the second, we show that the final sate might be a superposition of different states with different topologies; in the last, we apply the formalism to the case of homogeneous and isotropic cosmology. But first we describe the Fox algorithm to calculate the fundamental group of a branched-covering; we base our review on the papers \cite{fox1,fox2,fox3,fox4,fox5,fox6,fox7,dust2,comb}.

\subsection{The Fox algorithm}
\label{sec:fox}
Given a presentation for the fundamental group $\pi_1(S^3-\Gamma)$ of the complement of the branch locus $\Gamma$, one chooses an n-frame over the covering  with end-points $p^\prime_1 \cdots p_{n-1}^\prime$ where $n$ is the order of the covering  and identifies each of these points with the base point $p^\prime$ in $S^3-\Gamma$. The fundamental group of the resulting manifold is $\pi_1(M-p^{-1}(\Gamma))*F_{n-1}$, where $*$ indicates the free products and $F_{n-1}$ is a free group with $n-1$ elements. It can be shown that there is a homomorphism between $\pi_1(S^3-\Gamma)$ and $\pi_1(M-p^{-1}(\Gamma))*F_{n-1}$ which maps any word in the former into a word in the latter. A presentation of $\pi_1(M-p^{-1}(\Gamma))$ can be calculated from that of $\pi_1(S^3-\Gamma)$ as follows:
\begin{itemize}
\item if $\pi_1(S^3-\Gamma)$ has $e$ elements, the number of elements in $\pi_1(M-p^{-1}(\Gamma))*F_{n-1}$ is $n\,e$. For example, if the elements of $\pi_1(S^3-\Gamma)$ are $x$ and $y$ and the covering has 3 leaves, $\pi_1(M-p^{-1}(\Gamma))*F_{2}$ has 6 elements given by $x_0,x_1,x_2,y_0,y_1,y_2$.
\item if $\pi_1(S^3-\Gamma)$ has $r$ relations, the group $\pi_1(M-p^{-1}(\Gamma))*F_{n-1}$ has $n \, r$ relations. For example if $\pi_1(S^3-\Gamma)$ has one relation given by $xy$, and if it has the representations $x\rightarrow(0\,1)$ and $y\rightarrow(0\,1\,2)$ on $S_3$, there are three relations for $\pi_1(M-p^{-1}(\Gamma))*F_{2}$ given by:
\begin{equation}
x_0y_1\qquad x_1y_0 \qquad x_2y_2
\end{equation}
\item One has now to remove the free group $F_{n-1}$. This is done by identifying a Schreier tree (a set of word such that the final segment of a word is the initial segment of the following word) and by adding the relations to the others of $\pi_1(M-p^{-1}(\Gamma))*F_{n-1}$. In our example above a Schreier tree is given by $w_1=1,w_2=x,w_3=x^2$ and the relations one must add are $x_0=1$, $x_0x_1=1$. To these relations one must also add those coming from the cover representation; continuing with the above example, the relations to be added are $x_0x_1$ and $y_0y_1y_2$.  
\item Finally one has to work out all the consequences of the relations reducing the set of relations to the most simple one and so arrives at the presentation the group $\pi_1(M-p^{-1}(\Gamma))$.
\end{itemize}

\subsection{A change in topology due to the action of the Hamiltonian constraint}
\label{sec:change}
\begin{figure}
\centering
\subfigure{\includegraphics[scale=0.5]{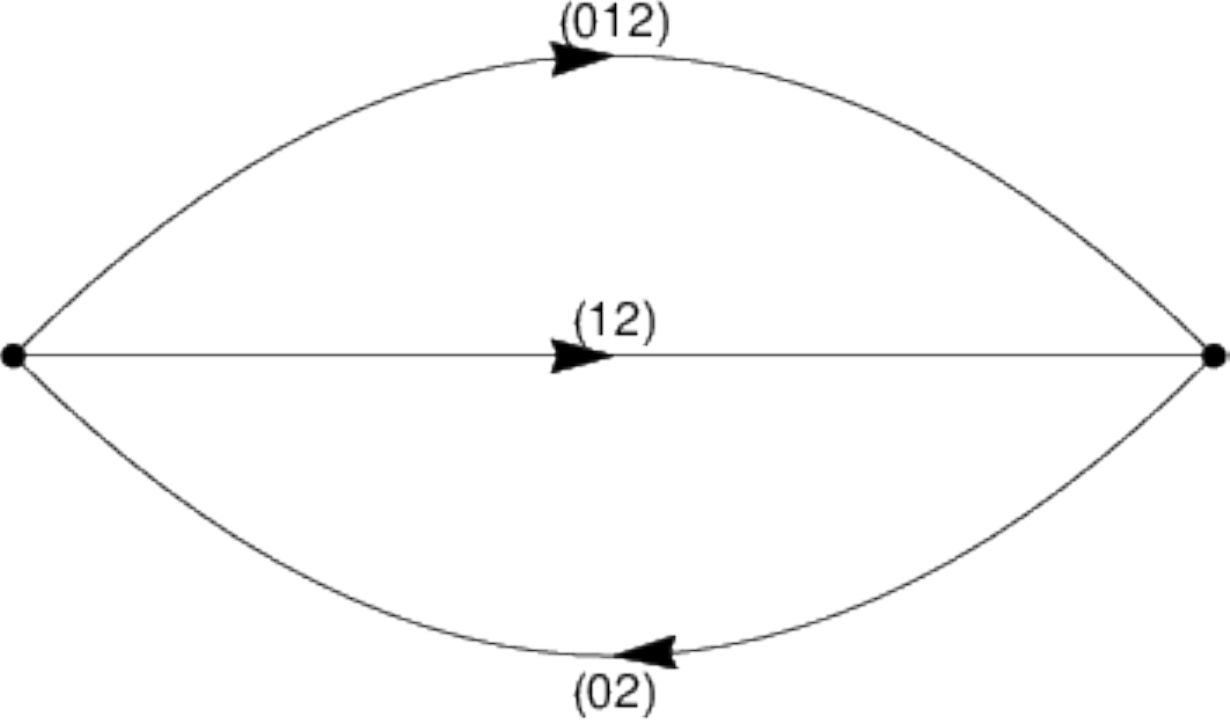}} \subfigure{\includegraphics[scale=0.5]{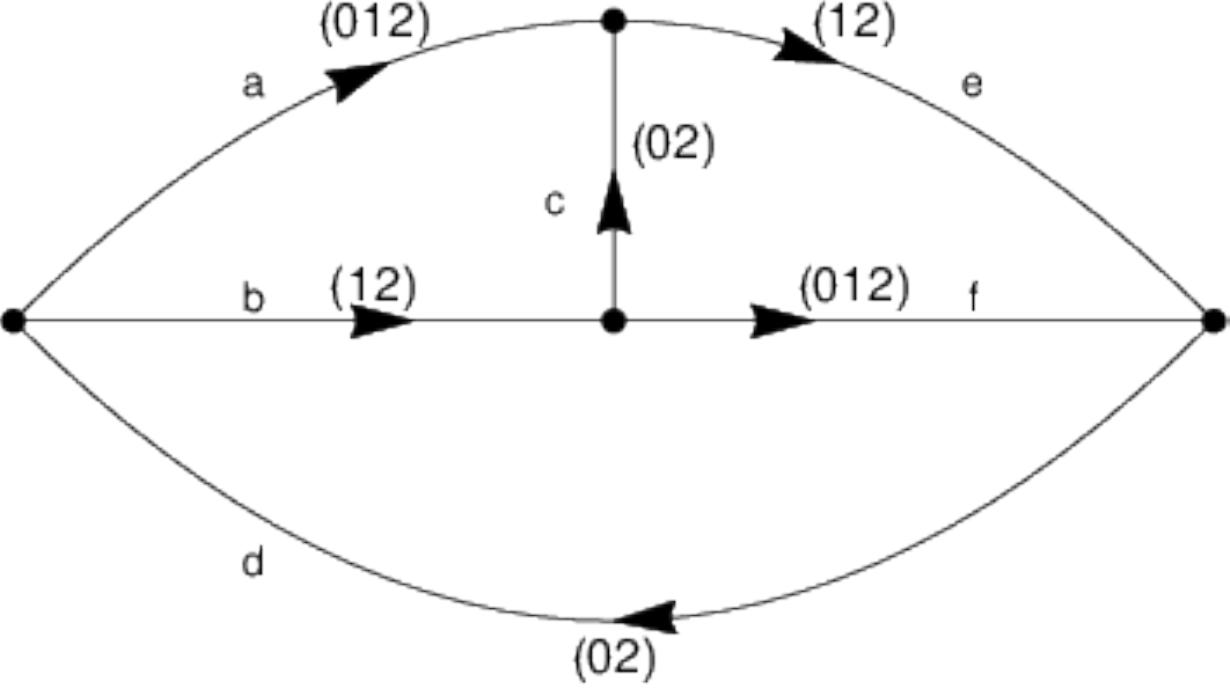}}
\caption{\emph{Left}: initial state. \emph{Right}: Graph produced by the action of the Hamiltonian constraint on the graph on the left.}\label{fig:change}
\end{figure}

For an example of a change in the topology of a manifold, consider the situation depicted in figure \ref{fig:change}: the Hamiltonian constraint acts on the graph on the left (call it $\Gamma_0$) producing that on the right ($\Gamma_1$); we shall now prove using Fox algorithm described above that the fundamental groups $\pi_1(M-p^{-1}(\Gamma_0))$ and $\pi_1(M-p^{-1}(\Gamma_1))$ are different: this means that the action of the Hamiltonian constraint has changed the topology of the manifold.

It was demonstrated in \cite{dust2} that for the graph in the left panel of figure \ref{fig:change} we have that $\pi_1(M-p^{-1}(\Gamma_0))$ is trivial; we therefore try to calculate the fundamental group of the graph in the right panel.

In the case of the right panel of figure \ref{fig:change}, a presentation of the group $\pi_1(S^3-\Gamma_1)$ is:
\begin{equation}\label{eq:fund1}
\pi_1(S^3-\Gamma_1)=\langle a,b,c,d,e,f; e=ac,ef=ab,b=fc\rangle
\end{equation}
where we have used the letters as reported in figure \ref{fig:change}. The presentation for the group $\pi_1(M-p^{-1}(\Gamma_1))*F_2$ is given by:
\begin{equation}
\pi_1(M-p^{-1}(\Gamma_1))*F_2 = \Big\langle \begin{array}{cccccc}
a_0 & b_0 & c_0 & d_0 & e_0 & f_0 \\
a_1 & b_1 & c_1 & d_1 & e_1 & f_1 \\
a_2 & b_2 & c_2 & d_2 & e_2 & f_2 \\
\end{array};\begin{array}{ccc}
e_0=a_0c_1 & e_0f_0=a_0b_1 & b_0=f_0c_1\\
e_1=a_1c_2 & e_1f_2=a_1b_2 & b_1=f_1c_2\\
e_2=a_2c_0 & e_2f_1=a_2b_0 & b_2=f_2c_0
\end{array} \Big\rangle
\end{equation}

The free group $F_2$ must be removed. We therefore consider the words $w_0=1,w_1=a,w_2=a^2$, which imply $a_0=a_1=1$. This simplifies greatly the relators, but we have also to add the conditions coming from the cover representation, which  are:
\begin{equation}
\begin{split}
&a_0a_1a_2=1  \quad b_1b_2=1=b_0 \quad c_0c_2=c_1=1 \\
&d_0d_2=1=d_1 \quad e_1e_2=1=e_0 \quad f_0f_1f_2=1
\end{split}
\end{equation}
which are read off from the graph colors. By considering all these relations and the fact that $a_0=a_1=1$, we find that the group of the manifold is given by:
\begin{equation}
\pi_1(M-p^{-1}(\Gamma_1))= \langle e_1, e_2; e_1=e_2^{-1}  \rangle \simeq \mathbb{Z}
\end{equation}

So we can see that the addition of the new edge led to a change of the fundamental group from the trivial group to one isomorphic to $\mathbb{Z}$.

\subsection{A superposition of topologies}
\label{sec:super}
\begin{figure}[ht]
\centering
\subfigure[]{\includegraphics[scale=0.3]{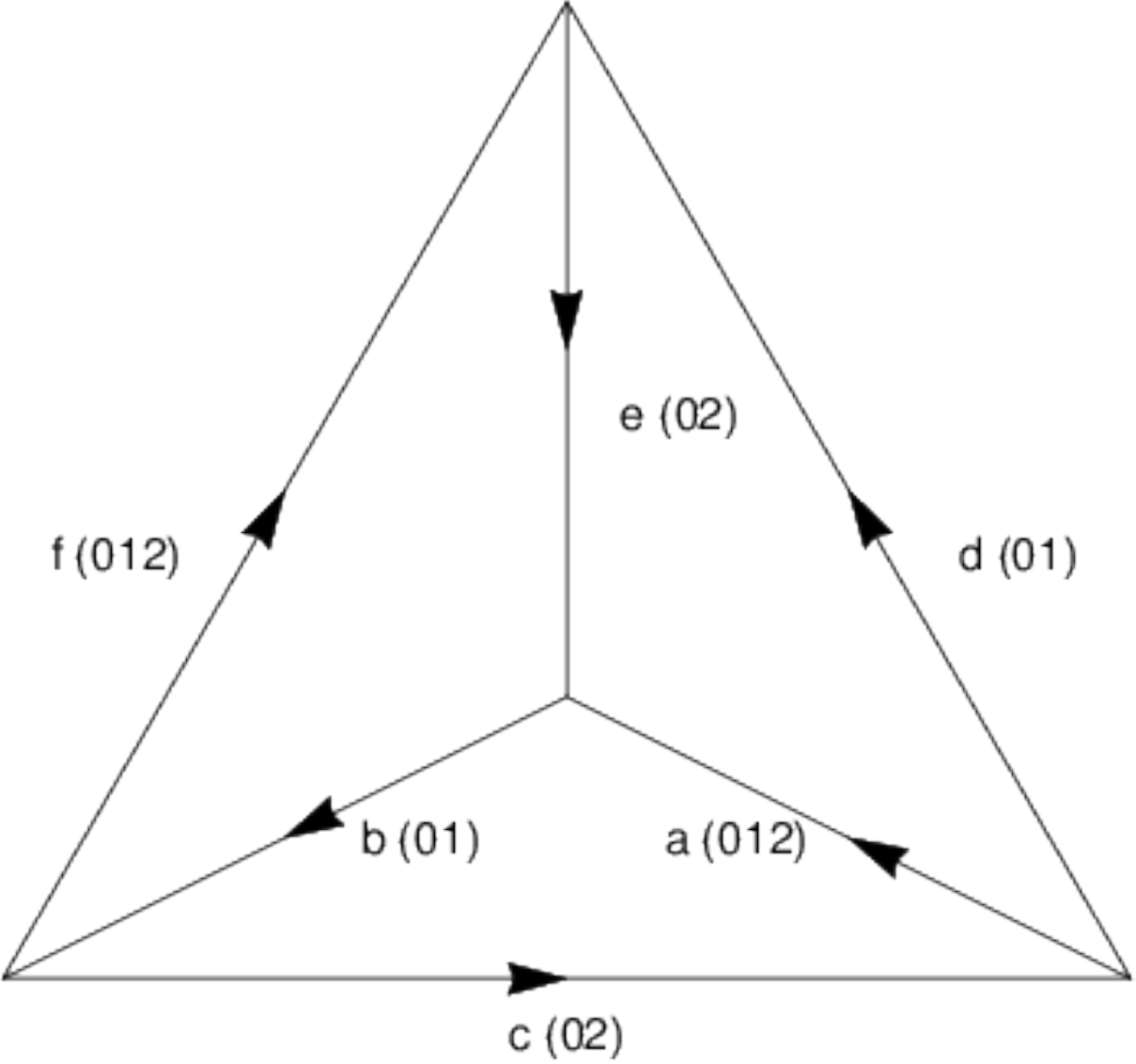}} \subfigure[]{\includegraphics[scale=0.3]{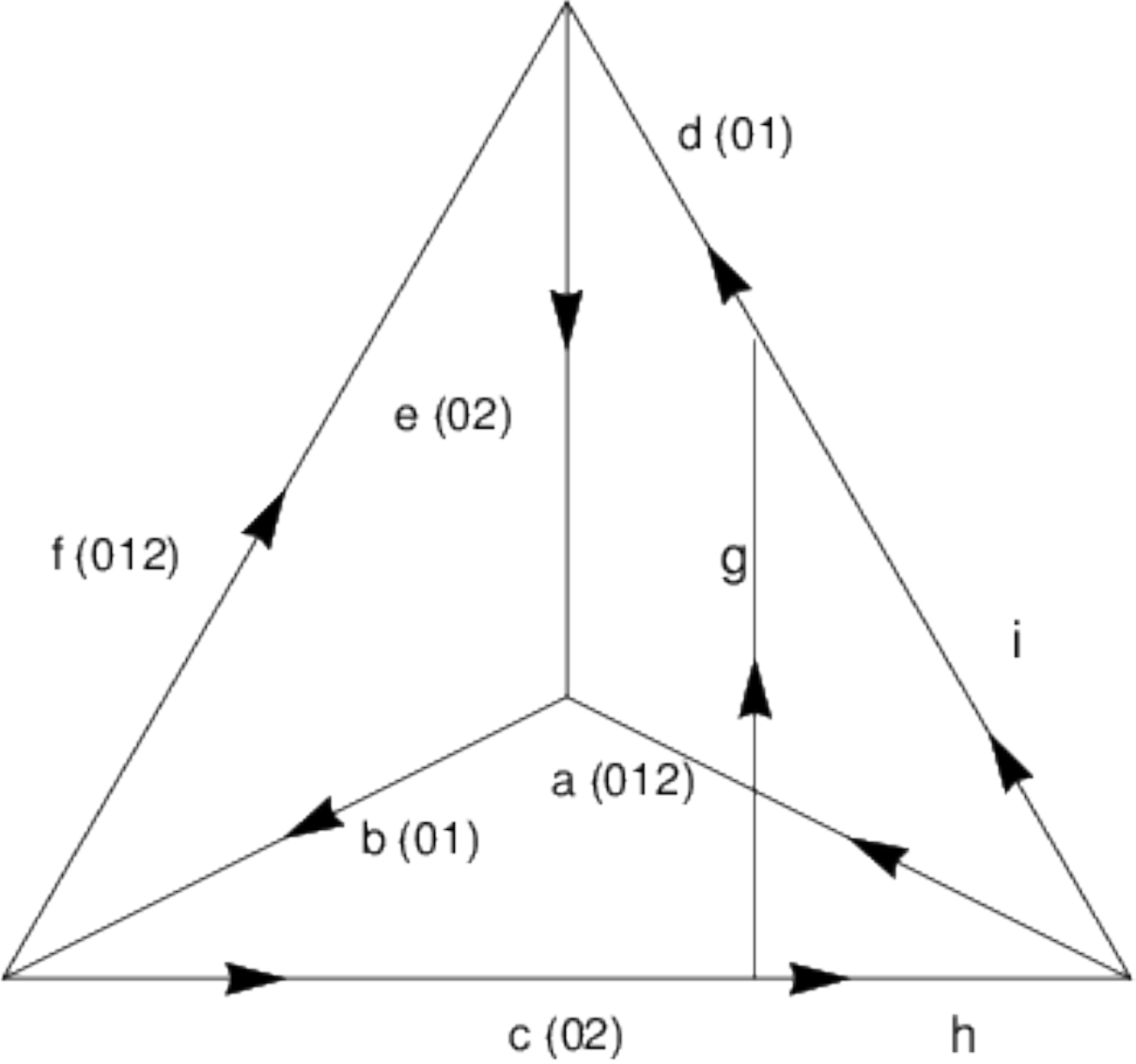}} \subfigure[]{\includegraphics[scale=0.3]{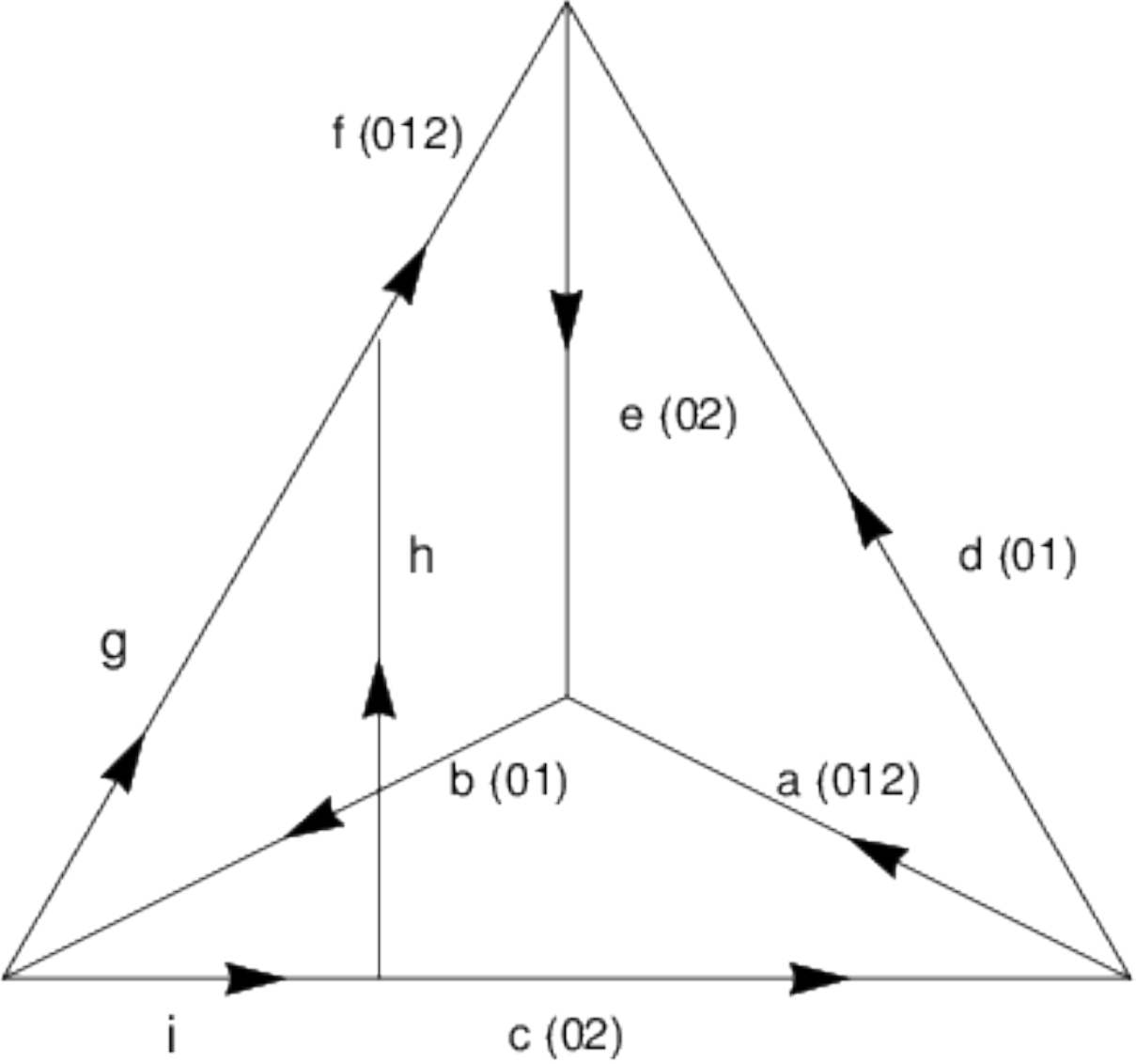}}
\subfigure[]{\includegraphics[scale=0.3]{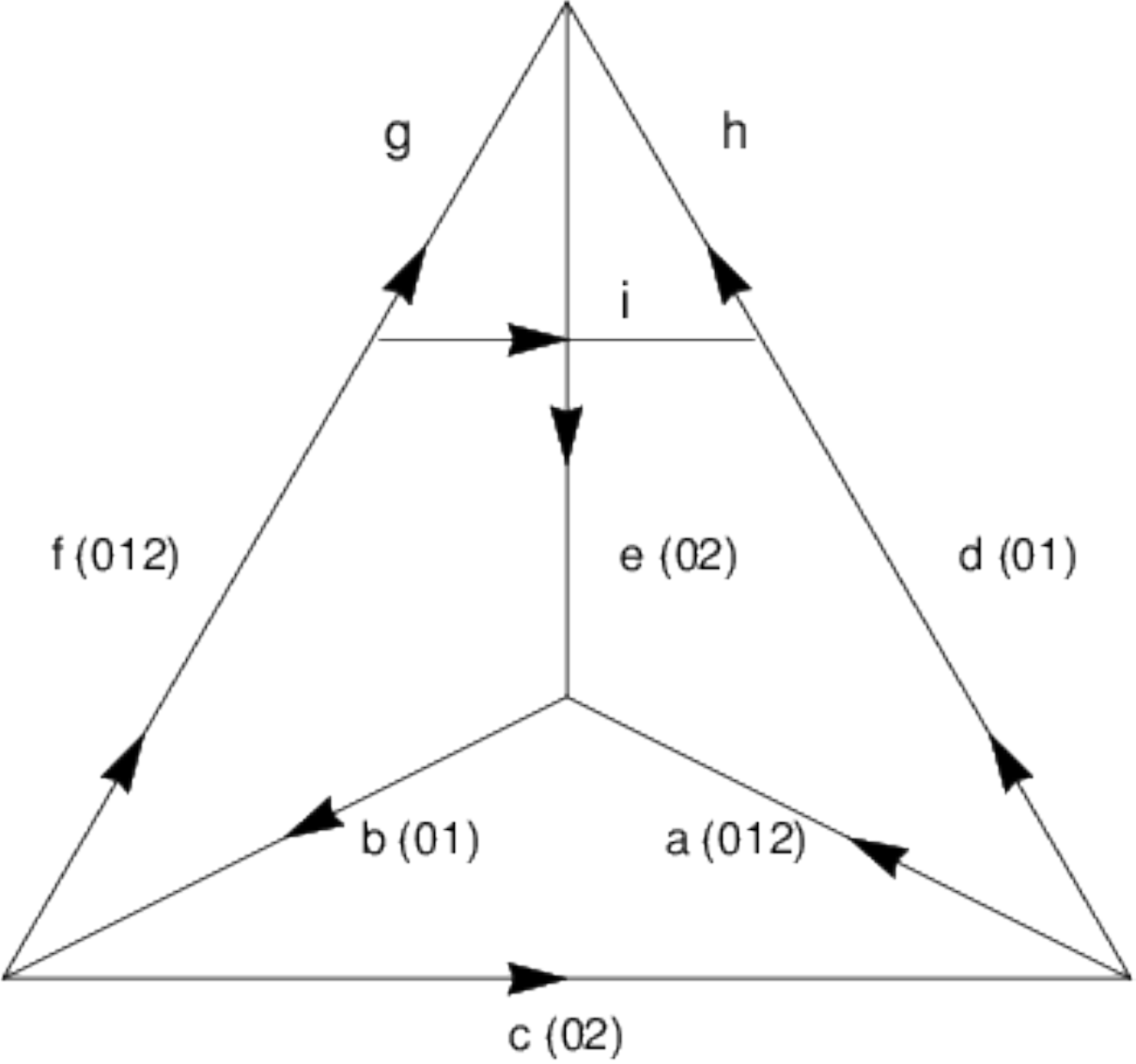}} \subfigure[]{\includegraphics[scale=0.3]{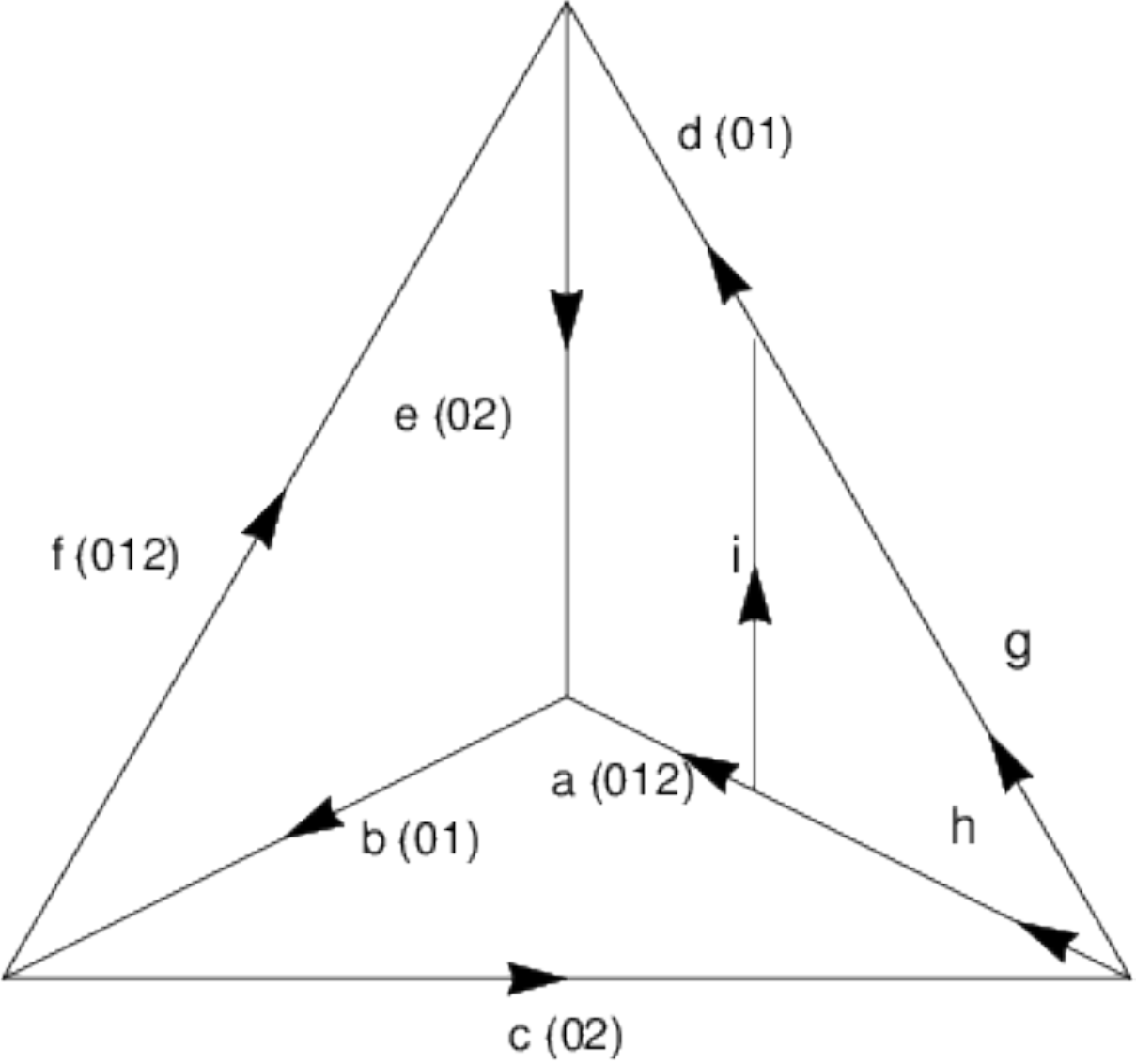}} \subfigure[]{\includegraphics[scale=0.3]{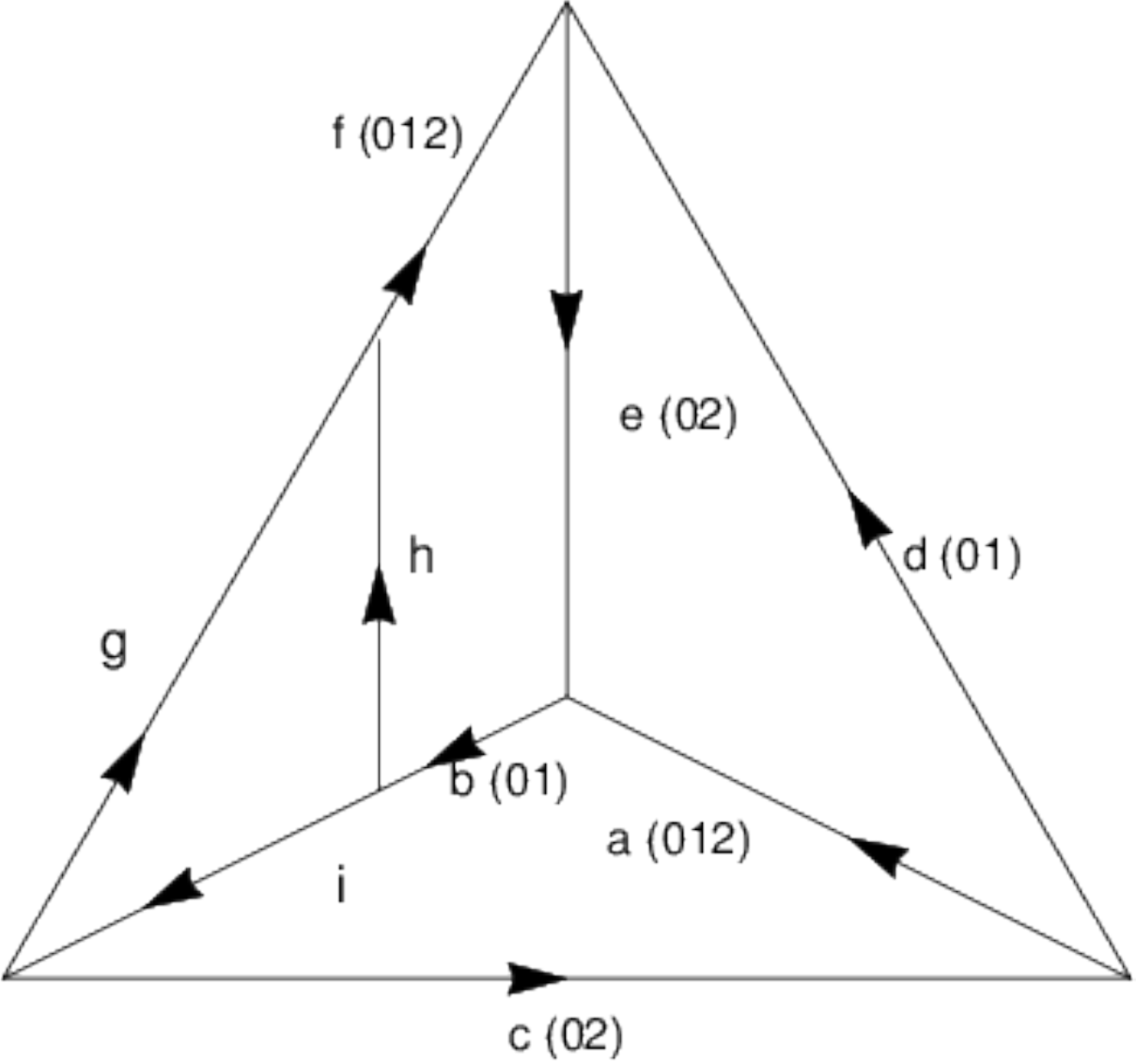}}
\subfigure[]{\includegraphics[scale=0.3]{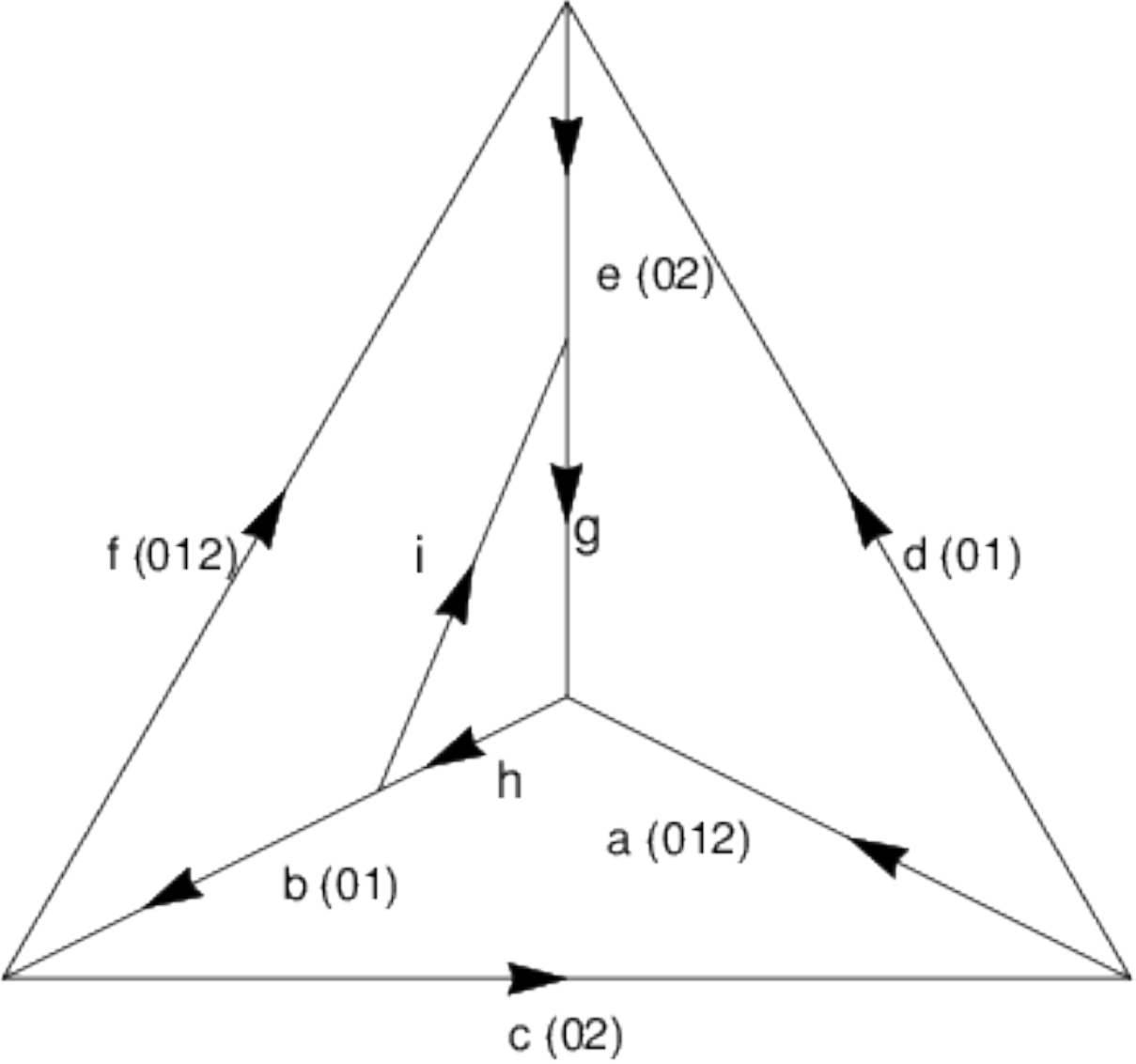}} \subfigure[]{\includegraphics[scale=0.3]{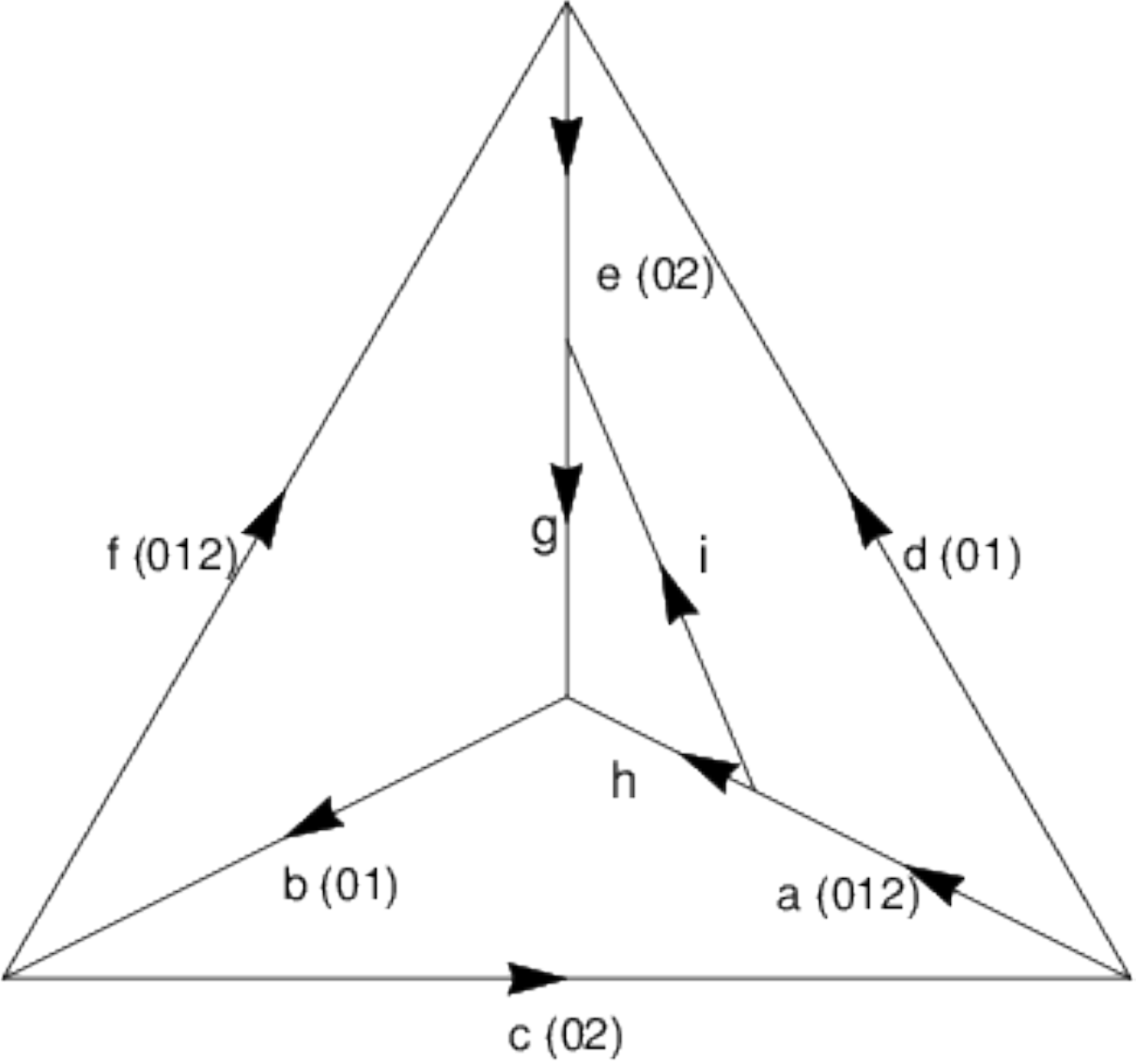}} \subfigure[]{\includegraphics[scale=0.3]{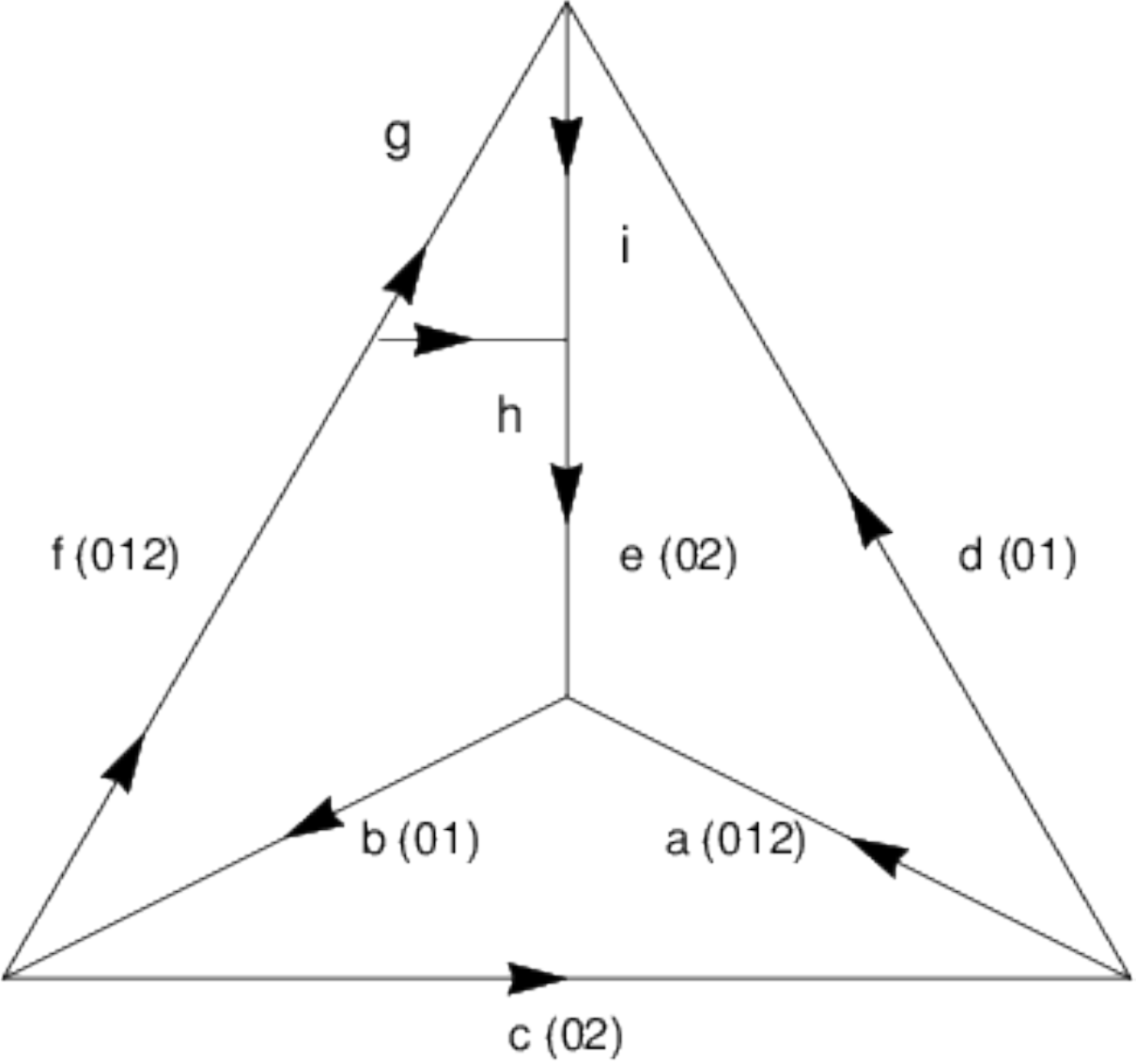}}
\subfigure[]{\includegraphics[scale=0.3]{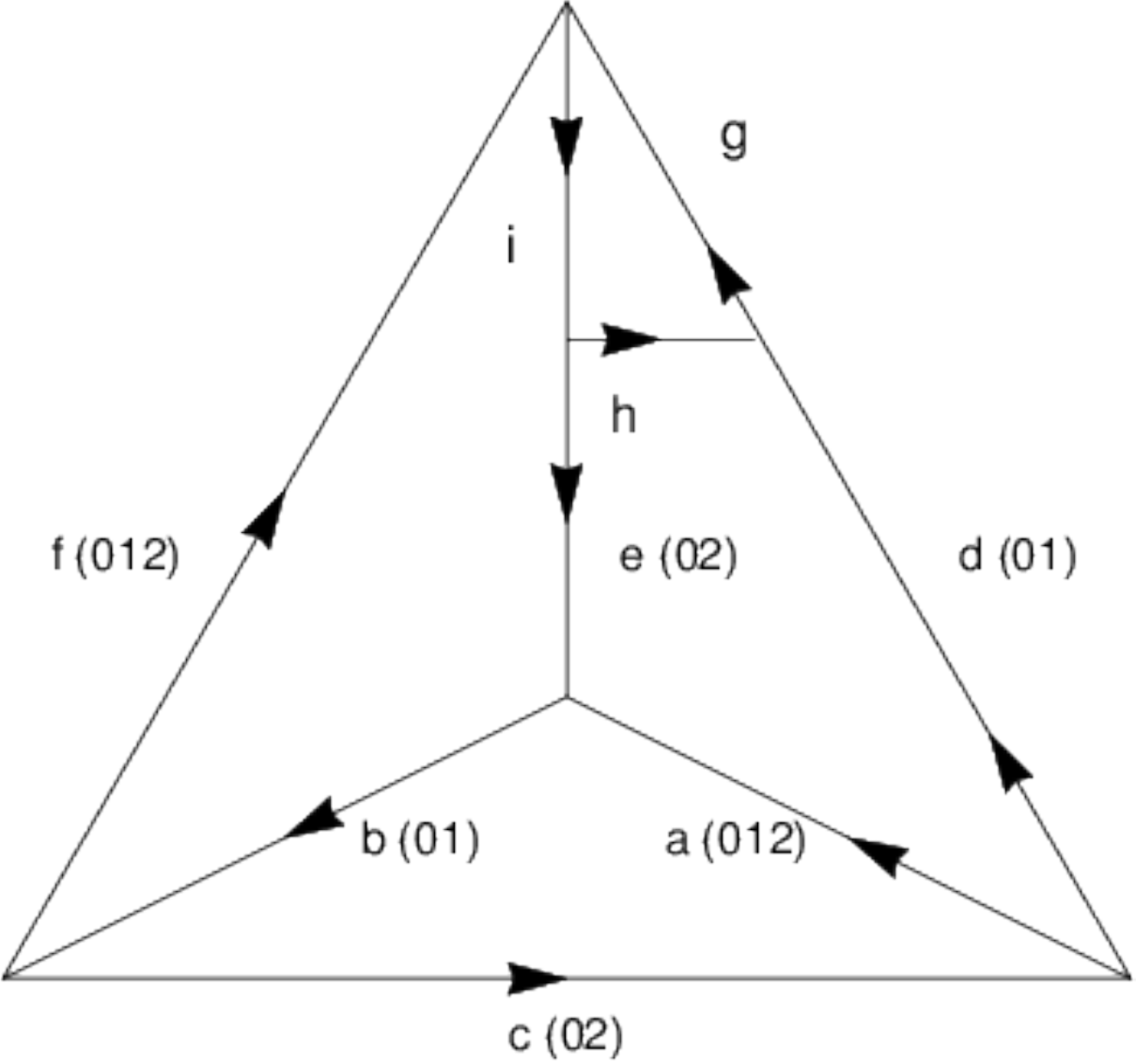}}
\caption{Tetrahedral graph and its modification due to the action of the Hamiltonian constraint.}\label{fig:tetra}
\end{figure}

We consider now the tetrahedral graph reported in figure \ref{fig:tetra}a along with its modifications due to the action of the Hamiltonian constraint (figures from 10b to 10j). By applying the formalism described in the subsection \ref{sec:fox}, we can calculate the fundamental group of the manifold associated to the graphs. In this case, however, there are multiple subcases for each graph. This is because the Wirtinger relations do not fix uniquely the various topological colors for the new edges. The results of the calculations for all the graphs are reported in table \ref{tab:risultati2}. As an example we report the calculation for the figure 10f.

The fundamental group of $S^3-\Gamma$ for the case 10f has presentation:
\begin{equation}
\pi_1(S^3-\Gamma)=\langle a,b,c,d,ef,g,h,i| g=fh, i=hb, i=cg, ae=b, b=cf, f=de \rangle
\end{equation}
The Wirtinger relations are:
\begin{equation}
\sigma_b=\sigma_i\sigma_h \qquad \sigma_g\sigma_h=\sigma_f \qquad \sigma_i=\sigma_g\sigma_c.
\end{equation}
where $\sigma_b,\sigma_c$ and $ \sigma_f$ are given. One can prove that there are two different solutions for the above relations as reported in table \ref{tab:risultati2}, which are:
\begin{equation}
\begin{array}{llll}
f1)&\sigma_i=(01) & \sigma_g=(012) & \sigma_h=\text{id}\\
f2)&\sigma_i=\text{id} & \sigma_g=(02) & \sigma_h=(01)
\end{array}
\end{equation}
where id means the identity. For the case $f1)$ the group $\pi_1(M-p^{-1}(\Gamma_1))*F_2$ has presentation:
\begin{equation}
\pi_1(M-p^{-1}(\Gamma_1))*F_2 = \Big\langle \begin{array}{ll}
a_0,b_0,c_0,d_0,e_0,f_0,g_0,h_0,i_0 &| g_0=f_0h_1, i_0=h_0b_0, i_0=c_0g_2, a_0e_1=b_0, b_0=c_0f_2, f_0=d_0e_1\\
a_1,b_1,c_1,d_1,e_1,f_1,g_1,h_1,i_1 &| g_1=f_1h_2, i_1=h_1b_1, i_1=c_1g_1, a_1e_2=b_1, b_1=c_1f_1, f_1=d_1e_0\\
a_2,b_2,c_2,d_2,e_2,f_2,g_2,h_2,i_2 &| g_2=f_2h_0, i_2=h_2b_2, i_2=c_2g_0, a_2e_0=b_2, b_2=c_2f_0, f_2=d_2e_2\\
\end{array} \Big\rangle
\end{equation}
As words for the Schreier tree we choose $w_1=1,w_2=a,w_3=a^2$, which give the relation $a_0=a_1=1$. The relations coming from the coloring of the graph are:
\begin{equation*}
\begin{array}{ccc}
a_0a_1a_2=1  & b_0b_1=1=b_2  & c_0c_2=1=c_1 \\
d_0d_1=1=d_2 & e_0e_2=1=e_1  & f_0f_1f_2=1 \\
g_0g_1g_2=1  & h_0=h_1=h_2=1 & i_0i_1=1=i_2
\end{array}
\end{equation*}
Now, with some algebra, one can reduce all the relations to the following presentation:
\begin{equation}
\pi_1(M) = \langle c_0; c_0^2 \rangle
\end{equation}
which is a presentation for the group $\mathbb{Z}_2$, therefore the manifold is a lens space $L(2,q)$ for some $q$.

For the case $f2)$ one has to proceed in the same way: the fundamental group $\pi_1(M-p^{-1}(\Gamma_1))*F_2$ has presentation:
\begin{equation}
\pi_1(M-p^{-1}(\Gamma_1))*F_2 = \Big\langle \begin{array}{ll}
a_0,b_0,c_0,d_0,e_0,f_0,g_0,h_0,i_0 &| g_0=f_0h_1, i_0=h_0b_2, i_0=c_0g_2, a_0e_1=b_0, b_0=c_0f_2, f_0=d_0e_1\\
a_1,b_1,c_1,d_1,e_1,f_1,g_1,h_1,i_1 &| g_1=f_1h_2, i_1=h_1b_1, i_1=c_1g_1, a_1e_2=b_1, b_1=c_1f_1, f_1=d_1e_0\\
a_2,b_2,c_2,d_2,e_2,f_2,g_2,h_2,i_2 &| g_2=f_2h_0, i_2=h_2b_0, i_2=c_2g_0, a_2e_0=b_2, b_2=c_2f_0, f_2=d_2e_2\\
\end{array} \Big\rangle
\end{equation}
as  words for the Schreier tree we choose again $w_1=1,w_2=a,w_3=a^2$, giving $a_0=a_1=1$. The relations coming from the coloring are:
\begin{equation*}
\begin{array}{ccc}
a_0a_1a_2=1  & b_0b_1=1=b_2  & c_0c_2=1=c_1 \\
d_0d_1=1=d_2 & e_0e_2=1=e_1  & f_0f_1f_2=1 \\
g_0g_2g_1=1  & h_0h_1=1=h_2  & i_0=i_1=i_2=1.
\end{array}
\end{equation*}
After some algebra, one finds that the fundamental group $\pi_1(M-p^{-1}(\Gamma))$ is trivial and therefore the manifold is $S^3$. 

We see that starting from a manifold ($S^3$ in this case) one can obtain different final manifolds after the action of the Hamiltonian constraint. This means that the final state of a system can be found in a superposition of topologies. From table \ref{tab:risultati2}, we see, however, that the action of the Hamiltonian constraint does not always lead to different topologies, in fact, it is the particular coloring of the initial graphs and the need to satisfy the Wirtinger relations that are responsible for the topology change.

\begin{table}[ht]
\centering
\begin{tabular}{cccc}
Graph & Permutations & Fundamental group & Manifold\\
\hline
a) && Trivial & $S^3$\\
\hline
\multirow{6}*{b)} & $\sigma_g=\text{id},\, \sigma_h=(02), \, \sigma_i=(01)$ & $\mathbb{Z}_2$ & Lens space $L(2,q)$\\
&$\sigma_g=(12),\, \sigma_h=(021), \, \sigma_i=(012)$ & $\mathbb{Z}_4$& Lens space $L(4,q)$\\
& $\sigma_g=(01),\, \sigma_h=(012), \, \sigma_i=\text{id}$ & Trivial & $S^3$\\
& $\sigma_g=(021),\, \sigma_h=(12), \, \sigma_i=(02)$ & Trivial & $S^3$\\
& $\sigma_g=(012),\, \sigma_h=(01), \, \sigma_i=(12)$ & Trivial & $S^3$\\
& $\sigma_g=(02),\, \sigma_h=\text{id}, \, \sigma_i=(021)$ & Trivial & $S^3$\\
\hline
\multirow{2}*{c)}& $\sigma_g=(02),\, \sigma_h=(01), \, \sigma_i=(012)$& $\langle c_0,g_0; g_0^2,g_0c_0g_0c_0^{-1} \rangle^{a}$ & \\
&$\sigma_g=\text{id},\, \sigma_h=(021), \, \sigma_i=(01)$ & Trivial & $S^3$\\
\hline
\multirow{2}*{d)}&$\sigma_g=(012),\, \sigma_h=(01), \, \sigma_i=\text{id}$&Trivial & $S^3$\\
&$\sigma_g=(02),\, \sigma_h=\text{id}, \, \sigma_i=(01)$ & Trivial & $S^3$\\
\hline
\multirow{3}*{e)}&$\sigma_g=(01),\, \sigma_h=(012), \, \sigma_i=(12)$& Trivial & $S^3$\\
&$\sigma_g=(02),\, \sigma_h=\text{id}, \, \sigma_i=(01)$ & Trivial & $S^3$\\
&$\sigma_g=(12),\, \sigma_h=(021), \, \sigma_i=(02)$ & Trivial & $S^3$\\
%
\hline
\multirow{2}*{f)} & $\sigma_g=(012),\, \sigma_h=\text{id}, \, \sigma_i=(01)$ & $\mathbb{Z}_2$& Lens space $L(2,q)$ \\
&$\sigma_g=(02),\, \sigma_h=(01), \, \sigma_i=\text{id}$ & Trivial & $S^3$\\
\hline
\multirow{3}*{g)}&$\sigma_g=(012),\, \sigma_h=(021), \, \sigma_i=(12)$& $\mathbb{Z}_4$& Lens space $L(4,q)$\\
& $\sigma_g=(021),\, \sigma_h=\text{id}, \, \sigma_i=(01)$ & $\mathbb{Z}_2$& Lens space $L(2,q)$\\
& $\sigma_g=\text{id},\, \sigma_h=(012), \, \sigma_i=(02)$ & $\mathbb{Z}_2$& Lens space $L(2,q)$\\
\hline
\multirow{3}*{h)} & $\sigma_g=\text{id},\, \sigma_h=(01), \, \sigma_i=(02)$ & $\mathbb{Z}_2$& Lens space $L(2,q)$\\
& $\sigma_g=(021),\, \sigma_h=(02), \, \sigma_i=(12)$ & $\mathbb{Z}_2$& Lens space $L(2,q)$\\
& $\sigma_g=(012),\, \sigma_h=(12), \, \sigma_i=(02)$ & $\mathbb{Z}_2$& Lens space $L(2,q)$\\
\hline
\multirow{3}*{i)} & $\sigma_g=(012),\, \sigma_h=\text{id}, \, \sigma_i=(02)$ & Trivial & $S^3$\\
& $\sigma_g=\text{id},\, \sigma_h=(021), \, \sigma_i=(01)$ & $\mathbb{Z}_2$& Lens space $L(2,q)$ \\
& $\sigma_g=(021),\, \sigma_h=(012), \, \sigma_i=\text{id}$ & Trivial & $S^3$\\
\hline
\multirow{3}*{j)} & $\sigma_g=(012),\, \sigma_h=(12), \, \sigma_i=(01)$ & Trivial & $S^3$ \\
& $\sigma_g=\text{id},\, \sigma_h=(01), \, \sigma_i=(02)$ & Trivial& $S^3$ \\
&$\sigma_g=(021),\, \sigma_h=(02), \, \sigma_i=(12)$ & Trivial &$S^3$\\
\hline
\end{tabular}
\caption{Fundamental groups of the manifolds represented by the graphs in figure \ref{fig:tetra}. \emph{id} means \emph{identity} and $\mathbb{Z}_n$ is the cyclic group of order $n$. \newline
$^{(a)}$ This group is the Higman-Neumann-Neumann extension of $\mathbb{Z}_2$ with $c_0$ as a stable letter, see \cite{comb}.}\label{tab:risultati2}
\end{table}

\subsection{Application to cosmology}
In this subsection we will apply the topspin network formalism to cosmology in the case in which topology is allowed to change;  we shall also calculate the transition amplitudes between the various final states.

\subsubsection{Variable topology}
If the topology is allowed to change, the basis of the system can be written as:
\begin{equation}
|\mu\rangle = |\mu_1\rangle\otimes|\mu_2\rangle\otimes|\mu_3\rangle = |\mu_1,\mu_2,\mu_3\rangle
\end{equation}
where $|\mu_i\rangle$ is the basis relative to the i-th cover and each $\langle c|\mu_i\rangle$ has the form \cite{bojo}:
\begin{equation}
\langle c|\mu_i\rangle = \exp\left( \dfrac{i \mu_i c}{2} \right).
\end{equation}
If the manifold is described as a branched covering, a curve for the calculation of the holonomy does not necessarily lie in one leaf of the cover, but in general will be in all of them; this means that there are three holonomies, one for each leaf of the cover, each with the form \cite{bojo}:
\begin{equation}
h_\gamma^{(i)} = \exp\left( \dfrac{i \delta c}{2} \right) \qquad (i)=\{(1),(2),(3)\}
\end{equation}
where $\delta$ is a function of $\mu$ \cite{ref}. Each of the $h_\gamma^{(i)}$ acts by shifting the index of the base state of the corresponding leaf of the cover, thus, for example:
\begin{equation}
h^{(1)}_\gamma(\delta) |\mu_1,\mu_2,\mu_3\rangle = |\mu_1+\delta,\mu_2,\mu_3\rangle 
\end{equation}
In this way, the action of the Hamiltonian constraint is given by a sum of terms:
\begin{equation}
\begin{split}
\mathcal{H}|\mu_1,\mu_2,\mu_3\rangle &\propto \dfrac{sgn(\mu_1)}{\delta_1^2} \left( V_{\mu_1+\delta_1,\mu_2,\mu_3} - V_{\mu_1-\delta_1,\mu_2,\mu_3} \right)\left( |\mu_1+4\delta_1,\mu_2,\mu_3 \rangle - 2 |\mu_1,\mu_2,\mu_3\rangle + |\mu_1-4\delta_1,\mu_2,\mu_3\rangle \right)+\\
&+\dfrac{sgn(\mu_2)}{\delta_2^2} \left( V_{\mu_1,\mu_2+\delta_2,\mu_3} - V_{\mu_1,\mu_2-\delta_2,\mu_3} \right)\left( |\mu_1,\mu_2+4\delta_2,\mu_3 \rangle - 2 |\mu_1,\mu_2,\mu_3\rangle + |\mu_1,\mu_2-4\delta_2,\mu_3\rangle \right)+\\
&+\dfrac{sgn(\mu_3)}{\delta_3^2} \left( V_{\mu_1,\mu_2,\mu_3+\delta_3} - V_{\mu_1,\mu_2,\mu_3-\delta_3} \right)\left( |\mu_1,\mu_2,\mu_3+4\delta_3 \rangle - 2 |\mu_1,\mu_2,\mu_3\rangle + |\mu_1,\mu_2,\mu_3-4\delta_3\rangle \right)=0
\end{split}
\end{equation}
where $V_{\mu_1,\mu_2,\mu_3}$ is the eigenvalue of the volume operator and is given by:
\begin{equation}
V|\mu_1,\mu_2,\mu_3\rangle =V_{\mu_1,\mu_2,\mu_3}|\mu_1,\mu_2,\mu_3\rangle=\left(\dfrac{4\pi \beta}{3} \ell_P^2\right)^{\frac{3}{2}} \left( \mu_1^{\frac{3}{2}}+\mu_2^{\frac{3}{2}}+\mu_3^{\frac{3}{2}} \right)|\mu_1,\mu_2,\mu_3\rangle
\end{equation}

The generic state of the system is $\Psi=\sum_{\mu_1,\mu_2,\mu_3}\psi_{\mu_1,\mu_2,\mu_3}|\mu_1,\mu_2,\mu_3\rangle$; the result of the action of  $\mathcal{H}$ on this state is again a difference equation:
\begin{equation}
\begin{split}
&\dfrac{sgn(\mu_1+4\delta_1)}{\delta_1^2} \left( V_{\mu_1+5\delta_1,\mu_2,\mu_3} - V_{\mu_1+3\delta_1,\mu_2,\mu_3} \right) \psi_{\mu_1+4\delta_1,\mu_2,\mu_3} -2 \dfrac{sgn(\mu_1)}{\delta_1^2} \left( V_{\mu_1+\delta_1,\mu_2,\mu_3} - V_{\mu_1-\delta_1,\mu_2,\mu_3} \right) \psi_{\mu_1,\mu_2,\mu_3} +\\
+& \dfrac{sgn(\mu_1-4\delta_1)}{\delta_1^2} \left( V_{\mu_1-3\delta_1,\mu_2,\mu_3} - V_{\mu_1-5\delta_1,\mu_2,\mu_3} \right) \psi_{\mu_1-4\delta_1,\mu_2,\mu_3} +\\
+&\dfrac{sgn(\mu_2+4\delta_2)}{\delta_2^2} \left( V_{\mu_1,\mu_2+5\delta_2,\mu_3} - V_{\mu_1,\mu_2+3\delta_2,\mu_3} \right) \psi_{\mu_1,\mu_2+4\delta_2,\mu_3} -2 \dfrac{sgn(\mu_2)}{\delta_2^2} \left( V_{\mu_1,\mu_2+\delta_2,\mu_3} - V_{\mu_1,\mu_2-\delta_2,\mu_3} \right) \psi_{\mu_1,\mu_2,\mu_3} +\\
+& \dfrac{sgn(\mu_2-4\delta_2)}{\delta_2^2} \left( V_{\mu_1,\mu_2-3\delta_2,\mu_3} - V_{\mu_1,\mu_2-5\delta_2,\mu_3} \right) \psi_{\mu_1,\mu_2-4\delta_2,\mu_3} +\\
+&\dfrac{sgn(\mu_3+4\delta_3)}{\delta_3^2} \left( V_{\mu_1,\mu_2,\mu_3+5\delta_3} - V_{\mu_1,\mu_2,\mu_3+3\delta_3} \right) \psi_{\mu_1,\mu_2,\mu_3+4\delta_3} -2 \dfrac{sgn(\mu_3)}{\delta_3^2} \left( V_{\mu_1,\mu_2,\mu_3+\delta_3} - V_{\mu_1,\mu_2,\mu_3-\delta_3} \right) \psi_{\mu_1,\mu_2,\mu_3} +\\
+& \dfrac{sgn(\mu_3-4\delta_3)}{\delta_3^2} \left( V_{\mu_1,\mu_2,\mu_3-3\delta_3} - V_{\mu_1,\mu_2,\mu_3-5\delta_3} \right) \psi_{\mu_1,\mu_2,\mu_3-4\delta_3} =0
\end{split}
\end{equation}
If there is matter, its Hamiltonian constraint must be added to the RHS of the above equations.

\begin{figure}[ht]
\centering
\subfigure{\includegraphics[scale=0.45]{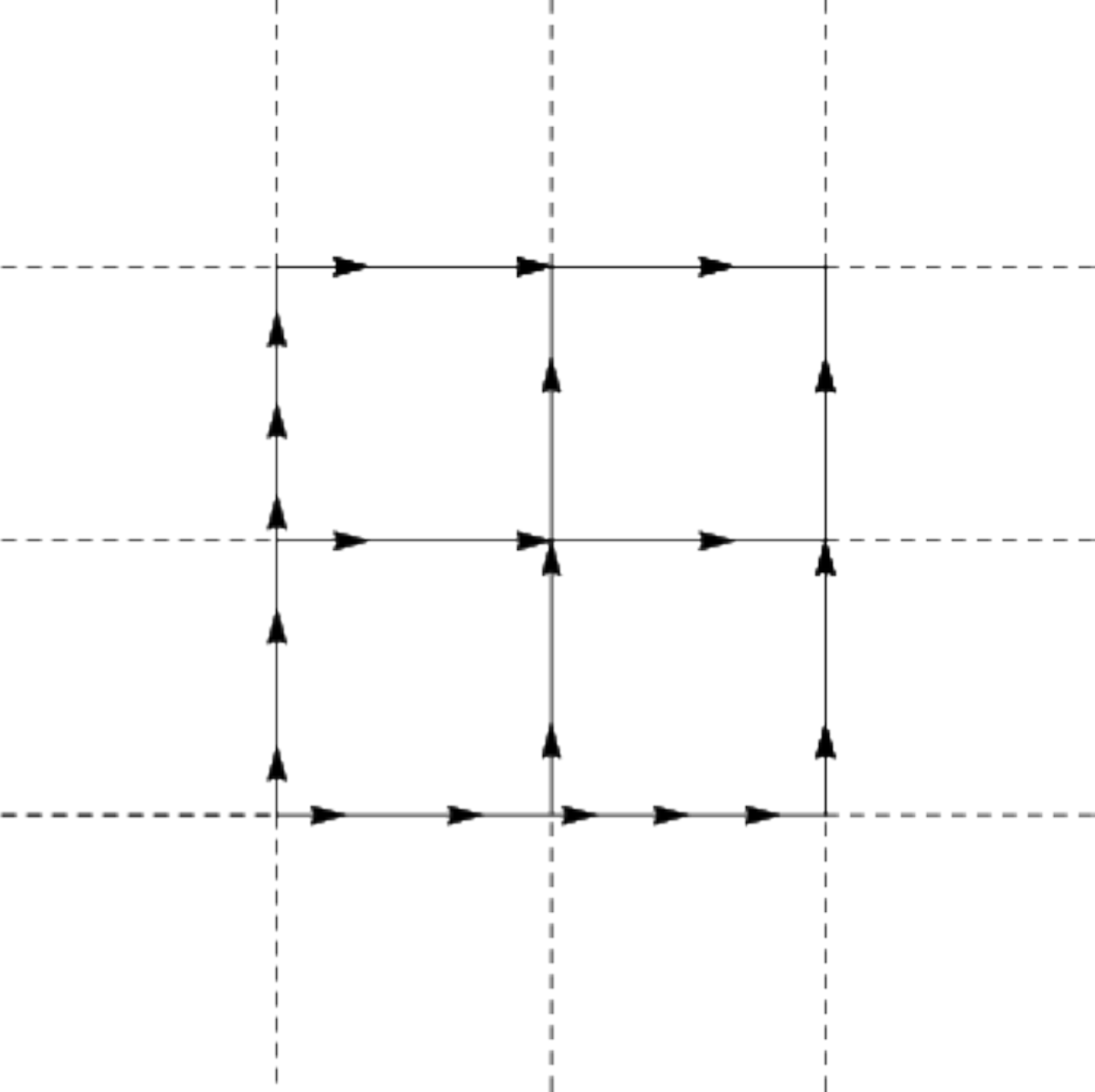}} \subfigure{\includegraphics[scale=0.45]{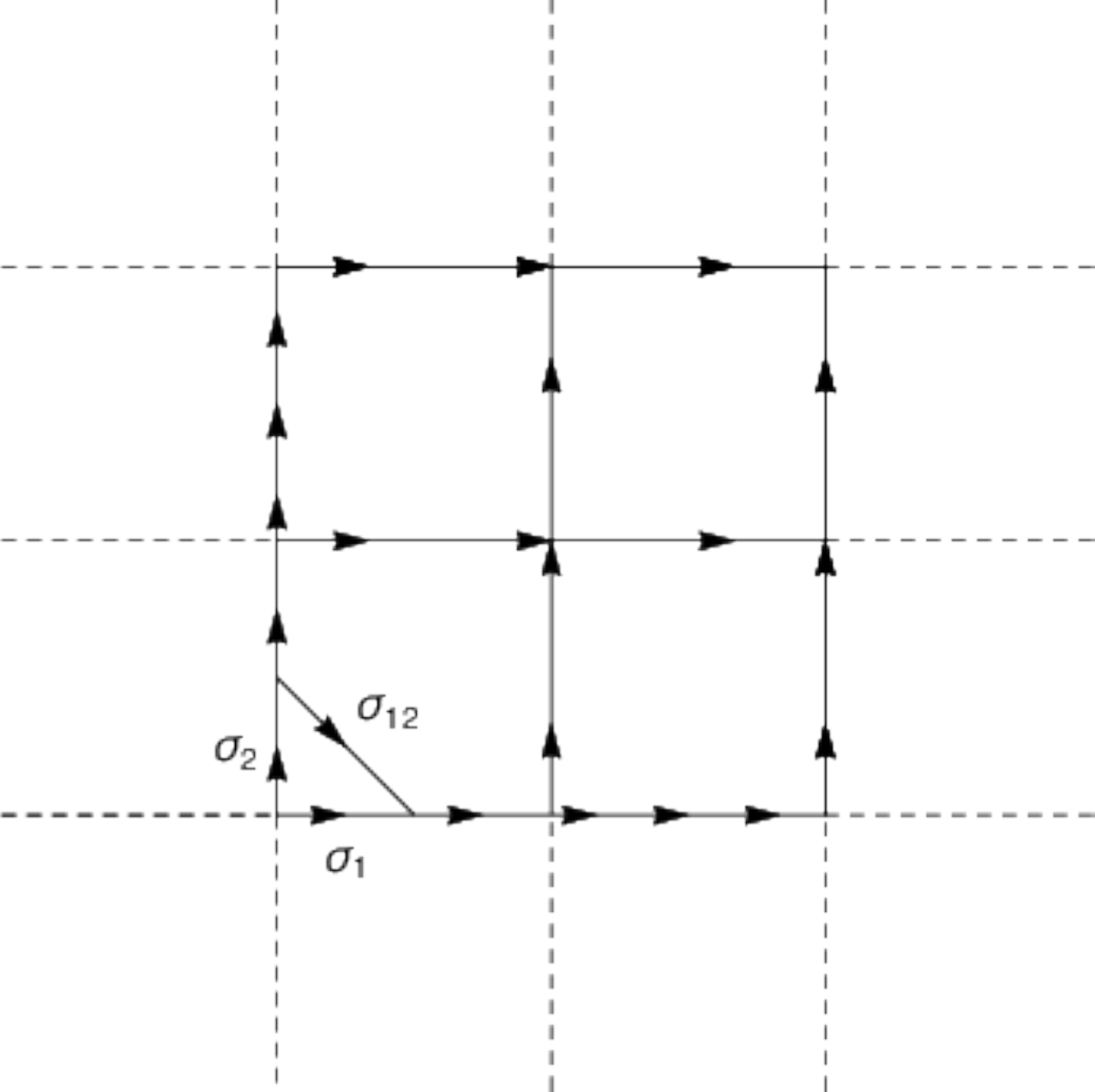}}
\caption{\emph{Left}: The graph for an homogeneous and isotropic manifold. \emph{Right}: the graph after the action of the Hamiltonian constraint.}\label{fig:latt}
\end{figure}

What about the topology? If we assume that the initial state is homogeneous and isotropic, as discussed in \cite{bojo}, one has to consider a graph  similar to  the lattice reported in the left panel of figure \ref{fig:latt}. The only closed, homogeneous and isotropic 3-manifold is $S^3$, whose fundamental group is trivial; given the graph in the left panel of figure \ref{fig:latt}, it is easy to prove that the only way one can get a trivial fundamental group is to impose that all the topological colors are trivial. In this case, the colors of the new edges $\sigma_{1},\sigma_{2}, \sigma_{12}$ created by the Hamiltonian constraint (for the notation see the right panel of figure \ref{fig:latt}) are undetermined, so there are 6 possibilities for $\sigma_{12}$, i.e.: $\sigma_{12}=\{\text{id}, (01),(02),(12),(012),(021)\}$; the corresponding colors for $\sigma_{1,2}$ are reported in table \ref{tab:isotropo} along with the fundamental groups of the new manifolds: we see that there is a superposition of three states and only one of them is $S^3$ again; the other possibilities are  the full torus (i.e. $T^2$ with the interior) and the lens space $L(3,q)$. 

\begin{table}[ht]
\centering
\begin{tabular}{cccccc}
Case&$\sigma_1$& $\sigma_2$ &$\sigma_{12}$ & $\pi_1$ & Manifold\\
\hline
(a) & id    & id    & id    & Trivial        & $S^3$\\
(b) & (01)  & (01)  & (01)  & $\mathbb{Z}$   & Full torus \\
(c) & (02)  & (02)  & (02)  & $\mathbb{Z}$   & Full torus \\
(d) & (12)  & (12)  & (12)  & $\mathbb{Z}$   & Full torus \\
(e) & (012) & (021) & (021) & $\mathbb{Z}_3$ & Lens space $L(3,q)$\\
(f) & (021) & (012) & (012) & $\mathbb{Z}_3$ & Lens space $L(3,q)$\\
\end{tabular}
\caption{The fundamental group of the new manifold created by the action of the Hamiltonian constraint on the graph of figure \ref{fig:latt} with the identity as topological color.}\label{tab:isotropo}
\end{table}

\subsubsection{Calculation of transition amplitudes and probabilities}
In this subsection we calculate the transition amplitudes and probabilities between the three final topologies of the cosmological model described above. 

If the transition amplitudes from the initial state $S^3$ to the final state $F_i$ are given by $W_{S^3\mapsto F_i}$ for $F_i=\{S^3,\text{torus}, L(3,q)\}$, the transition probabilities are:
\begin{equation}
P_{S^3\mapsto F_i} = \dfrac{|W_{S^3\mapsto F_i}|^2}{|W_{S^3\mapsto S^3}|^2+|W_{S^3\mapsto \text{torus}}|^2+|W_{S^3\mapsto L(3,q)}|^2}.
\end{equation}
In order to include topological colors in the calculation, one has to calculate the transition amplitude $W_{S^3\mapsto F_i}$ for each color combination in table \ref{tab:isotropo}.

\begin{figure}[ht]
\centering
\includegraphics[scale=0.7]{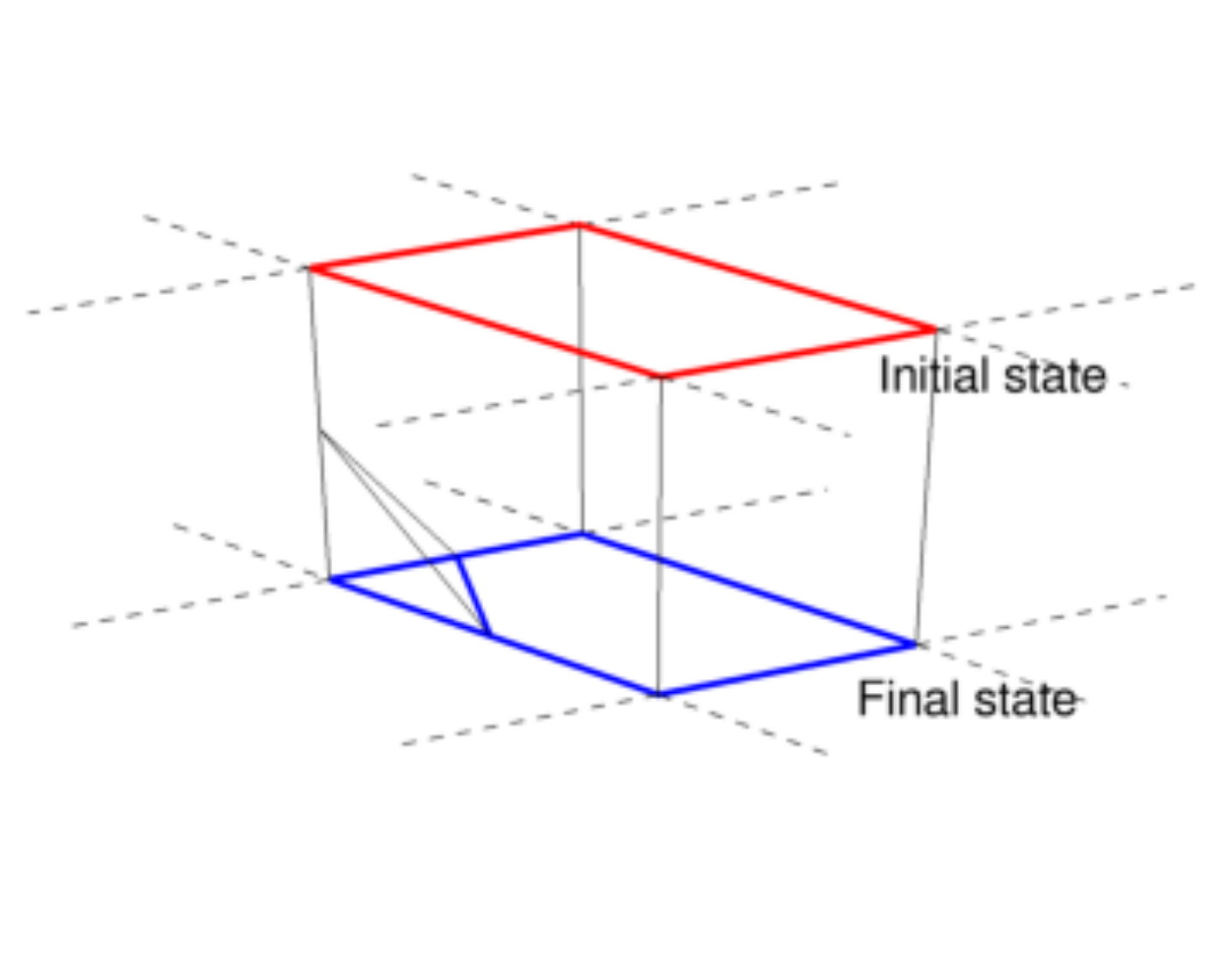}
\caption{Spin foam used in the calculation of the transition amplitude. Only the thick part must be considered in the calculation. The red links must be considered for the calculation of the initial state $S^3$, while the blue part must be considered for the calculation of the final state $F_i$.}\label{fig:sf}
\end{figure}

With a moment of reflection, one can deduce that in order to calculate the transition amplitudes, one has to focus on the thick part of the spinfoam reported in figure \ref{fig:sf}.
 
Following \cite{rov,sf}, the transition amplitude can be calculated using the formula:
\begin{equation}
W_{S^3\mapsto F_i} = \int_{SU(2)} dh\, \mathcal{W}_{S^3\mapsto F_i}(h)\psi_{S^3\mapsto F_i}(h)
\end{equation}
where $\psi_{S^3\mapsto F_i}(h)$ is the state of the system with $h_l\in SU(2)$ and
\begin{equation}
\mathcal{W}_{S^3\mapsto F_i}(h)=\sum_{\text{colorings}}\int_{SL(2,\mathbb{C})} dg\, \prod_f P_f(g,g^\prime,h_l).
\end{equation}
where the sum runs over all the topological colorings and $ P_f(g,g^\prime,h_l)$ will be defined shortly. Following \cite{sf}, it can be calculated that the coherent state $\psi_{S^3\mapsto F_i}(h)$ is given by:
\begin{equation}
\psi_{S^3\mapsto F_i}(h)= \psi_{S^3}(h)\psi_{F_i}(h)
\end{equation}
where we have defined, following \cite{sf} (see also \cite{coher1,coher2,coher3}):
\begin{equation}
\psi_{s}(h)=\prod_\ell \left( \sum_j d_j \exp\left( \dfrac{-j(j+1)}{2\sigma} \right) Tr_j\left[ D^j(n^{-1}h^{-1}n^\prime)D^j(e^{\frac{z\sigma_3}{2}}) \right] \right)
\end{equation}
where $d_j=2j+1$ and $\ell$ are the links, while $n$ is a $SU(2)$ element given by 
\begin{equation}
n=\exp(-i\phi \sigma_3/2)\exp(i\theta \sigma_2/2)\exp(i\phi \sigma_3/2).
\end{equation}
 When $s=S^3$ one has to consider the red part of figure \ref{fig:sf}, while if $s=F_i$ one has to consider the blue part. In the above equation $z$ is a complex number, while $\sigma$ gives the peakedness of the state. In order to have a state peaked both in the area and in the extrinsic curvature, one has to impose \cite{sf}:
\begin{equation}\label{eq:sigma}
\sigma = \sqrt{j_0}
\end{equation} 
where $j_0$ is the value of the angular momentum that minimizes the quantity
\begin{equation}
\dfrac{j(j+1)}{2\sigma} - Re(z)j
\end{equation}
which is $j_0=Re(z)\sigma$, thus from equation \eqref{eq:sigma}, we find $Re(z)=\sqrt{j_0}$. The imaginary part of $z$ is $\zeta$, the extrinsic curvature of the manifold \cite{sf}.

$\mathcal{W}_{S^3\mapsto F_i}(h)$ can be splitted in a way similar the the state $\psi_{S^3\mapsto F_i}(h)$; in fact, defining:
\begin{equation}
P^{S^3}(g,g^\prime,h_l)= \prod_{red}\left( \sum_j d_j D^{\beta j,j}_{j,m,l,p}(g^\prime) D^{\beta j,j}_{l,p,j,n}(g^{-1}) D^j_{m,n}(h_{red}) \right)
\end{equation}
which is calculated with the red part of figure \ref{fig:sf}, and 
\begin{equation}
P^{F_i}(g,g^\prime,h_l)= \prod_{blue}\left( \sum_j d_j D^{\beta j,j}_{j,m,l,p}(g^\prime) D^{\beta j,j}_{l,p,j,n}(g^{-1}) D^j_{m,n}(h_{blue}) \right)
\end{equation}
which is calculated with the blue part, we have:
\begin{equation}
\mathcal{W}_{S^3\mapsto F_i}(h_l)=\sum_{\text{coloring}}\int_{SL(2,\mathbb{C})} dg\, P^{S^3}(g,g^\prime,h_l) P^{F_i}(g,g^\prime,h_l).
\end{equation}

Putting it all together and performing the integration over $SU(2)$, we find:
\begin{equation}\label{eq:fin}
W_{S^3\mapsto F_i} = \sum_{\text{coloring}}\int_{SL(2,\mathbb{C})} dg\, \prod_\ell \left( \sum_j d_j \exp\left( -\dfrac{j(j+1)}{2\sigma} \right) D^{\beta j,j}_{j,m,l,p}(g^\prime)D^{\beta j,j}_{l,p,j,n}(g^{-1}) D^j_{m,n}(n^\prime e^{\frac{z\sigma_3}{2}} n^{-1}) \right).
\end{equation}
where $\ell$ are the links, both blue and red.

Following \cite{sf}, the D-matrices $D^j_{m,n}(e^{\frac{z\sigma}{2}})$ are dominated by the contribution of the higher-valued angular momenta, and can be substituted by:
\begin{equation}\label{eq:prima}
D^j_{m,n}(e^{\frac{z\sigma}{2}}) \mapsto \delta_m^j \delta_n^j e^{zj}.
\end{equation}

The integration in $SL(2,\mathbb{C})$ can be performed by substituting it with two $SU(2)$ integrations and one over a parameter $r$ as follows \cite{sf}:
\begin{equation}\label{eq:integrale}
\int_{SL(2,\mathbb{C})} dg \longmapsto \int_0^\infty dr \dfrac{\sinh^2r}{4\pi} \int_{SU(2)}du \int_{SU(2)} dv.
\end{equation}
Similarly, the Wigner matrices $D^{\beta j,j}_{j,m,l,p}$ can be substituted with Wigner-D matrices:
\begin{equation}\label{eq:seconda}
D^{\beta j,j}_{j,m,l,n}= D^j_{m,p}(u) d_{j,l,p}(r) D^l_{p,n}(v^{-1})
\end{equation}
where \cite{sf}:
\begin{equation}
d_{j,l,p}(r)=\sqrt{d_j}\sqrt{d_l}\int_0^1 dt\left[ d^l_{jp}\left( \dfrac{t e^{-r}-(1-t)e^r}{t e^{-r}+(1-t)e^r} \right) d^j_{jp}(2t-1)(t e^r+(1-t)e^r)^{i\beta j-1}\right]
\end{equation}
and $d^j_{lp}(\cos\delta)$ are Wigner $SU(2)$ matrices, while $\beta$ is the Barbero-Immirzi constant. Finally, Wigner D-matrices have the following property \cite{wigner}:
\begin{equation}
D^j_{m,n}(U)=(-1)^{m-n}\left[ D^j_{-m,-n}(U) \right]^*
\end{equation}
where the star indicates complex conjugation; with this and the fact that the D-matrices satisfy the orthogonality condition \cite{wigner}:
\begin{equation}
\int_{SU(2)}dU D^{j_1}_{m_1,n_1}(U)\left[D^{j_2}_{m_2,n_2}(U)\right]^* = \dfrac{8\pi^2}{2j_1+1} \delta_{j_1,j_2}\delta_{m_1,m_2}\delta_{n_1,n_2},
\end{equation}
we find that the integral of the product of two D-matrices is given by:
\begin{equation}\label{eq:terza}
\int_{SU(2)}dU D^{j_1}_{m_1,n_1}(U)D^{j_2}_{m_2,n_2}(U) = \dfrac{8\pi^2}{2j_1+1}(-1)^{m_2-n_2}\delta_{j_1,j_2}\delta_{m_1,-m_2}\delta_{n_1,-n_2}.
\end{equation}

Substituting all the $SL(2,\mathbb{C})$ integrations in \eqref{eq:fin} with $SU(2)$ integrations as in \eqref{eq:integrale}, substituting formulae \eqref{eq:prima} and    \eqref{eq:seconda} in the result and performing all of the integrations over $u$ and $v$ using equation \eqref{eq:terza}, we find:
\begin{equation}
W_{S^3\mapsto F_i} =\sum_{\text{coloring}} \prod_\ell \sum_{j_\ell} \left[ (2j_\ell+1)^{-1} \exp\left( -\dfrac{j_\ell(j_\ell+1)}{2\sigma} + z j_\ell\right)\,I_{\ell ,j_\ell} \right]
\end{equation}
where we have defined:
\begin{equation}
I_{\ell,j_\ell}=16\pi^3 \sum_{n_\ell}\int^\infty_0  dr_\ell \, \sinh^2r_\ell\, \Big( d_{j_\ell,j_\ell,n_\ell}(r_\ell) \otimes d_{j_\ell,j_\ell,-n_\ell}(r_\ell) \Big).
\end{equation}


For the transition $S^3\mapsto S^3$, all the links have trivial colors and we are in the usual situation of LQG. One can find the semiclassical state by using saddle-points method \cite{rov}; noting that all the eleven links give the same contribution, one finds that the transition amplitude is given by:
\begin{equation}
W_{S^3\mapsto S^3} \propto \exp\left( -11\dfrac{(i+2z\sigma)^2}{8\sigma} \right)
\end{equation}
The classical limit is obtained by the rapidly oscillating phase \cite{rov}:
\begin{equation}
W_{S^3\mapsto S^3} \propto \exp\left( -i\dfrac{11}{8} j_0 \zeta_{S^3} \right)
\end{equation}
where $\zeta_{S^3}$ is the extrinsic curvature of the hypersphere.

For the case of the transition $S^3\mapsto $torus, one has to deal with links of different colors: eight (four links in the initial state and four in the final state) have again trivial colors, while the other three may be colored (01),(01),(01) or (02),(02),(02) or (12),(12),(12), therefore there will be three different contributions to the amplitude:
\begin{equation}
\begin{split}
W_{S^3\mapsto \text{torus}} &\propto \exp\left( -\dfrac{i}{2} j_0\zeta_{S^3} \right)\exp\left( -\dfrac{i}{2} j_0 \zeta_{T} \right) \Bigg[ \exp\left( -i\dfrac{3}{8}j_0\zeta_T \right)\Bigg|_{(01)}+\\
&+\exp\left( -i\dfrac{3}{8} j_0\zeta_T \right)\Bigg|_{(02)}+\exp\left( -i\dfrac{3}{8} j_0 \zeta_T \right)\Bigg|_{(12)} \Bigg]=\\
W_{S^3\mapsto \text{torus}}&\propto 3\exp\left[ -i j_0 \left( \dfrac{\zeta_{S^3}}{2} + \dfrac{7\zeta_{T}}{8} \right) \right]
\end{split}
\end{equation}
where $\zeta_T$ is the extrinsic curvature of the torus; in the above equation, we have indicated with a vertical bar on the right of the summation term the topological color of the link.

Finally, the transition amplitude $S^3\mapsto L(3,q)$ is given by:
\begin{equation}
\begin{split}
W_{S^3\mapsto L(3,q)} &\propto \exp\left( -\dfrac{i}{2} j_0\zeta_{S^3} \right)\exp\left( -\dfrac{i}{2} j_0\zeta_L \right) \Bigg[ \exp\left( -i\dfrac{1}{8} j_0 \zeta_L \right)\Bigg|_{(012)}\exp\left( -i\dfrac{1}{4} j_0 \zeta_L \right)\Bigg|_{(021)}+\\
&+\exp\left( -i\dfrac{1}{8} j_0 \zeta_T \right)\Bigg|_{(021)}\exp\left( -i\dfrac{1}{4} j_0 \zeta_L \right)\Bigg|_{(012)} \Bigg]=\\
W_{S^3\mapsto L(3,q)} &\propto 2 \exp\left[ -i j_0 \left( \dfrac{\zeta_{S^3}}{2} + \dfrac{7\zeta_L}{8} \right)\right]
\end{split}
\end{equation}
where $\zeta_L$ is the extrinsic curvature of the lens space.

\section{Conclusion}
\label{sec:concl}
Motivated by the fact that in quantum gravity, quantum mechanics might lead to the evolution of the topology of the manifold, we have used the formalism of topspin network in the context of LQG to characterize the topological state of a system. We have described a method based on Fox algorithm to calculate the fundamental group of the manifold and applied this formalism to three toy models: with the first, we have shown that the topology can actually change after the action of the Hamiltonian constraint; in the second, we have found that the final state of the system might be a superposition of states with different topologies;  finally, we have applied the formalism to a homogeneous and isotropic cosmological model calculating the difference equation that describes the state of the system and finding that the final system is in a superposition of three topologies: $S^3$, the full torus or a lens space; for this last case we have also calculated the transition amplitudes and transition probabilities from $S^3$ to the three final states. 

For the cosmology application, we notice that even if observations favor the flat Friedmann-Lema\^itre-Robertson-Walker (FLRW) model \cite{planck_sat,planck_sat2}, these observations are limited to the local part of the observable Universe and, moreover, while the three FLRW models are derived with the requirement of homogeneity and (local) isotropy, this does not mean that the spatial slices are necessarily $S^3$, $\mathbb{R}^3$ (i.e. flat) or $H^3$ (i.e. hyperbolic with negative curvature), these models are only the three simplest possibilities and the global topology of the Universe might be different from that implied by these models and more exotic, see for example the review \cite{cosmo}. In fact $S^3,\mathbb{R}^3$ and $H^3$ are the universal cover of any homogeneous and (locally) isotropic Universe; this means that any other space can be constructed by topological operations such as gluing together or identifying some parts of these manifolds. One could hope to \emph{see} the signature of the topology of the Universe with observations: for example one could hope to see the light from the same source coming from different directions in the case of a toroidal topology; the absence (up to now) of these particular signatures leads to a lower bound on the volume of the Universe and, by themselves, do not rule out exotic topologies. Changes in the topology in the history of the Universe might be observable in a similar way, for example hidden in the cosmic microwave background, or in changes in the propagation of light from distance sources, or, if they happened in the early Universe, they might be observable in the gravitational waves background.

As a consequence of the three examples considered above, we see that the change of the topology of the manifold is a general phenomenon due to the particular color configuration of the initial graph and to the Wirtinger relations that do not always give a unique result for the colors of the extraordinary edges added by the action of the Hamiltonian constraint. This shows the need to extend the LQG approach to include in the theory the modification of the manifold topology during the evolution of the system under study. Situations more general than the simple ones considered in this paper will lead to the same conclusion that the topology will evolve in time along with the geometry of the manifold, stressing this need to extend LQG, since more complicated graphs will lead to more possible combinations of colors and topologies.

\end{document}